\documentclass[aip,jcp,amsmath,amssymb,reprint,floatfix]{revtex4-2}

\usepackage[utf8]{inputenc}
\usepackage{graphicx}
\usepackage{mathtools}
\usepackage{booktabs}
\usepackage{hyperref}
\usepackage{csquotes}
\usepackage{color}
\usepackage{braket}
\usepackage{subcaption}
\hypersetup{hidelinks}
\graphicspath{{./figures/}}

\newcommand{\JWmaps}{\xmapsto[\scriptstyle{\mathrm{JW}}]{}}
\newcommand{\Paritymaps}{\xmapsto[\scriptstyle{\Pi}]{}}
\newcommand{\RotatedParitymaps}{\xmapsto[\scriptstyle{\Pi_R}]{}}
\newcommand{\JWParitymaps}{\xmapsto[\scriptstyle{\Pi \, \circ \, \mathrm{JW}}]{}}

\begin{document}

 \title{Nonlinear Canonical Transformations for Studying Spin-\texorpdfstring{$1/2$}{1/2} Hamiltonians}


\author{Swarnamoy Ghosh}
\email[Author to whom correspondence should be addressed: ]{swarnamoy.ghosh@rice.edu}
\affiliation{Department of Chemistry, Rice University, Houston, Texas 77005, USA}

\author{Thomas M. Henderson}
\affiliation{Department of Chemistry, Rice University, Houston, Texas 77005, USA}
\affiliation{Department of Physics and Astronomy, Rice University, Houston, Texas 77005, USA}

\author{Gustavo E. Scuseria}
\email[Author to whom correspondence should be addressed: ]{Gustavo.E.Scuseria@rice.edu}
\affiliation{Department of Chemistry, Rice University, Houston, Texas 77005, USA}
\affiliation{Department of Physics and Astronomy, Rice University, Houston, Texas 77005, USA}

\begin{abstract}
Strongly correlated quantum systems, including frustrated spin lattices, remain challenging for conventional coupled-cluster methods because their ground states often exhibit strong multireference character and symmetry breaking. Although multireference and symmetry-projected coupled-cluster theories offer promising avenues, they are not yet black-box methods and often require substantial system-specific development. Spin Hamiltonians such as the XXZ and $J_1$-$J_2$ models provide compact but demanding benchmarks for these challenges. In particular, spin-BCS-based coupled-cluster methods may require high excitation ranks to recover the exact ground-state physics in strongly correlated regimes. In this work, we introduce a nonlinear spin--fermion--spin (SFS) transformation designed to map a strongly correlated spin Hamiltonian into a new spin Hamiltonian whose ground state is more accurately described at the mean-field level. For
a representative SFS transformation, we identify a compact Clifford circuit that maps transformed product states to circuit-dressed, generally entangled states in the original representation. The transformed Hamiltonian can then be treated using broken-symmetry mean-field theory and post-mean-field methods. Benchmarks for the XXZ and $J_1$--$J_2$ Hamiltonians show that spin mean-field theory in the transformed representation reduces the energy error relative to conventional spin-BCS mean field. The spin-BCS-based coupled-cluster calculations for the one-dimensional XXZ model provide substantially improved accuracy relative to the corresponding untransformed methods. In 2D, the accuracy of the transformed mean-field solutions depends on the site ordering and boundary conditions because nonlocal strings appear. These results show that exact transformations of operator representations can
serve as many-body preconditioners for mean-field and low-rank
coupled-cluster approximations to strongly correlated spin
systems. 
\end{abstract}

\maketitle

\section{Introduction}
Spin-\(\frac{1}{2}\) Hamiltonians offer a compact route to studying strongly correlated quantum phenomena across several areas of physics and chemistry. They have been used to describe magnetic materials, high-temperature superconductivity, and chemically relevant magnetic processes such as nitrogen fixation.~\cite{Heisenberg1928,Anderson1987,Tazhigulov2022} Spin models like the XXZ and $J_1-J_2$ Heisenberg Hamiltonians have gained considerable attention in the last couple of decades and exhibit magnetic order, frustration, and strong quantum fluctuations. However, except in a few special cases~\cite{Kitaev2006}, spin lattice models in more than one dimension are not exactly solvable, making approximate many-body methods essential. At the same time, these models remain challenging because their low-energy states can be strongly correlated, particularly near critical points or in frustrated regimes.\cite{Amico2008Entanglement,Eisert2010AreaLaws, Schollwoeck2011DMRG}.

In recent work \cite{Henderson2022StrongWeak,Henderson2023EJW}, it has been shown that the difficulty of a many-body problem depends on the algebraic representation in which it is expressed. For example, the Jordan-Wigner (JW) transformation \cite{JordanWigner1928} converts a Hamiltonian built from $\mathfrak{su}(2)$ operators into a fermionic Hamiltonian, and that fermionic Hamiltonian is usually less strongly correlated and can be solved accurately with mean-field methods. Although the mapped Hamiltonian becomes many-body due to JW strings, nevertheless, matrix elements can be evaluated at polynomial cost using nonorthogonal Wick's theorem. \cite{Chen2023RobustWick,Balian1969NonunitaryBogoliubov} It can be easily shown that in the JW mapped one-dimensional XXZ model, the JW strings drop out and at $\Delta=0$, which is the most strongly correlated regime in the spin representation, the fermionic Hamiltonian is non-interacting and therefore can be solved exactly.

JW transformation is only one of the members of a broader family of transformations connecting spin and fermionic degrees of freedom.\cite{JordanWigner1928,Whitfield2011QuantumSimulation,Havlicek2017OperatorLocality,Chien2026SimulatingFermions} Parity,\cite{Seeley2012BK} Bravyi--Kitaev,\cite{BravyiKitaev2002} and related encodings distribute occupation and parity information differently among the spin variables. In the quantum information community, these transformations are often used to map fermions to qubits. 

Standard mean-field theories usually generate reference states through linear canonical transformations of the underlying operators. In fermionic systems, for example, Hartree-Fock (HF) and Hartree-Fock-Bogoliubov (HFB) \cite{RingSchuck1980} theories optimize over linear transformations of creation and annihilation operators while preserving the canonical anticommutation relations. This provides a powerful and tractable class of wavefunctions, but it explores only a restricted portion of the space of possible representations.

A broader possibility is to perform a nonlinear canonical transformation of the Hamiltonian before applying mean-field theory, so that the transformed Hamiltonian is exactly equivalent to the original one and has the same spectrum. However, approximate methods are not invariant under such transformations. Thus, a nonlinear change of representation may convert a strongly correlated problem into one whose ground state is better described by a simple mean-field or low-rank post-mean-field ansatz. In this work, 'nonlinear' refers to the action of the transformation on the local on-site operator: an on-site spin operator may be mapped to a product or string of operators, thereby changing its locality and body rank. The induced transformation $(A\mapsto UAU^\dagger)$ on the full operator space remains linear and unitary.



In this work, we explore a different route to solve spin-$\frac{1}{2}$ Hamiltonians by performing a canonical transformation that is constructed by sequentially mapping spin-$\frac{1}{2}$ operators to a new set of spin-$\frac{1}{2}$ operators using JW and parity transformations. Although the JW and parity mappings are non-canonical transformations within a fixed operator algebra, they are exact and  invertible mappings between spin and fermionic representations and thus their composition yields a canonical transformation of
the spin algebra. We first map spin operators to fermionic operators using the JW transformation and then map the fermionic operators back to a new set of spin variables using a parity transformation. The result is what we call the spin-fermion-spin (SFS) transformation. The final Hamiltonian is again a spin Hamiltonian, but with a different operator rank. The idea is simply that nonlinear canonical transformations are not generally obvious to derive, but chaining a faithful spin to fermion mapping with a faithful fermion to spin mapping provides a canonical transformation between spins, and this transformation will generally change the body rank and locality of the elementary spin operators.


Our goal is to determine whether these nonlinear SFS canonical transformations provide a better starting point for approximate many-body calculations on spin Hamiltonians. By comparing results in the original and transformed representations, we assess whether the transformed frame weakens the correlation, allowing the exact ground state to be captured more accurately within a mean-field manifold. In this sense, this work extends the idea of spin-to-fermion strong--weak duality to a spin-to-spin canonical transformation framework suitable for post-mean-field methods.


\section{Spin-Fermion Transformations}
\label{Spin-Fermion Transformations}
A spin-1/2 degree of freedom and a spinless fermion both give rise to a two-dimensional Hilbert space. For spin 1/2, the Hilbert space is spanned by the two states $\ket{\downarrow}$ and $\ket{\uparrow}$, whereas the Hilbert space of a single spinless fermion is spanned by the states $\ket{0}$ and $\ket{1}$ in the occupation number representation.  This makes it possible to map spin-1/2 states onto spinless fermion states and vice versa. The real challenge is to extend these ideas to map between states of $M$ fermions and states of $M$ spins, since operators acting on different spins commute, while operators acting on different fermions anticommute. A useful spin-fermion transformation must preserve this structure. This is accomplished by introducing nonlocal string operators, which compensate for the difference between commuting spin operators and anticommuting fermionic operators. The Jordan--Wigner transformation is the earliest example of such a mapping, and as we have said, many related transformations, including parity and Bravyi--Kitaev encodings, have since been developed. Here, we want to summarize those that we consider in this work.

\subsection{Operator Algebras}
Before deriving the SFS transformation, we first introduce the operator notation for spin, fermion, and transformed spin representations. The bare fermionic creation and annihilation operators for level $p$ are denoted as $c_{p}^{\dagger}$ and $c_p$ respectively, and they act on the basis states as
\begin{subequations}
\begin{align}
c_{p}^{\dagger} \, \ket{0_p} &= \ket{1_p},  &   c_{p}^{\dagger} \ket{1_p} &= 0,
\\
c_{p} \, \ket{1_p} &= \ket{0_p},            &   c_{p} \ket{0_p} &= 0.
\end{align}
\end{subequations}
The number operator $n_p$, defined as
\begin{equation}
n_p = c^{\dagger}_p \, c_p,
\end{equation}
determines whether the level $p$ is occupied or empty, and the fermionic states $\ket{0_p}$ and $\ket{1_p}$ are eigenstates of $n_p$ with eigenvalues 0 and 1, respectively. The operators obey several algebraic relations:
\begin{subequations}
\begin{align}
\{c_p^\dagger,c_p\} &= c_p^\dagger \, c_p + c_p \, c_p^\dagger = 1,
\\
c_p^\dagger \, c_p^\dagger &= c_p \, c_p = 0,
\\
[c_p^\dagger,c_p] &= c_p^\dagger \, c_p - c_p \, c_p^\dagger = 2 \, n_p - 1,
\\
[n_p,c_p^\dagger] &= c_p^\dagger,
\\
[n_p,c_p] &= -c_p.
\end{align}
\end{subequations}

Similarly, we can define spin raising and lowering operators
\begin{subequations}
\begin{align}
\label{spin-1/2 canonical}
S_p^+ \, \ket{\downarrow_p} &= \ket{\uparrow_p},    &   S_p^+ \, \ket{\uparrow_p} &= 0,
\\
S_p^- \, \ket{\uparrow_p} &= \ket{\downarrow_p},    &   S_p^- \, \ket{\downarrow_p} &= 0.
\end{align}
\end{subequations}
The spin-1/2 states $\ket{\downarrow_p}$ and $\ket{\uparrow_p}$ are eigenstates of the operator $S_p^z$ with eigenvalues $-1/2$ and $+1/2$. Moreover, we have similar algebraic relations:
\begin{subequations}
\begin{align}
\{S_p^+,S_p^-\} &= 1,
\\
S_p^+ \, S_p^+ &= S_p^- \, S_p^- = 0,
\\
[S_p^+,S_p^-] &= 2 \, S_p^z,
\\
[S_p^z,S_p^+] &= S_p^+,
\\
[S_p^z,S_p^-] &= -S_p^-.
\end{align}
\end{subequations}

For a single site, then, $S_p^+$ and $c_p^\dagger$ play the same role, as do $S_p^-$ and $c_p$; $S_p^z$ plays the same role as $n_p - 1/2$, and the algebras of $\{S_p^+,S_p^-,S_p^z\}$ and of $\{c_p^\dagger,c_p,n_p - \frac{1}{2}\}$ are the same.

Differences arise when we have multiple sites. Fermionic operators on different sites anticommute:
\begin{subequations}
\begin{align}
\{c_{p}, c_{q}\} &= 0,
\\
\{c_{p}, c_q^{\dagger}\} &= \delta_{pq},
\end{align}
\end{subequations}
while the spin operators on different sites commute:
\begin{subequations}
\begin{align}
[S^+_p,S^-_q] &= 2 \, \delta_{pq} \, S^z_p,
\\
[S^z_p,S^{\pm}_q] &= \pm\delta_{pq} \, S^{\pm}_p.
\end{align}
\end{subequations}

For this work, it is more convenient to work with Majorana operators and spin-1/2 Pauli operators. We define Majorana operators $\gamma_{1,p}$ and $\gamma_{2,p}$ as
\begin{subequations}
\begin{align}
\gamma_{1,p} &= c_p + c_p^{\dagger},
\\
\gamma_{2,p} &= \text{i}\,(c_p -c^{\dagger}_p).
\end{align}
\end{subequations}
These Majoranas satisfy the anticommutation relations
\begin{equation}
\{\gamma_{\mu,p},\gamma_{\nu,q}\} = 2 \, \delta_{\mu \nu} \, \delta_{pq}
\end{equation}
and are Hermitian and involutions:
\begin{subequations}
\begin{align}
\gamma_{\mu,p}^{\dagger} &= \gamma_{\mu,p},
\\
\gamma_{\mu,p}^2 &= 1.
\end{align}
\end{subequations}
We can show that
\begin{equation}
\text{i}\,\gamma_{1,p}\,\gamma_{2,p} = 1 -2 \, n_p.
\end{equation}

To define Pauli operators, we first need to introduce $S_p^x$ and $S_p^y$, given by
\begin{equation}
S^{\pm}_p = S_p^x \pm \mathrm{i} \, S^y_p.
\end{equation}
We have
\begin{equation}
[S^{i}_p,S^{j}_q] = \mathrm{i} \, \delta_{pq} \, \epsilon_{ijk} \, S^k_p,
\end{equation}
where $i,j,k \in \{x,y,z\}$. Then the Pauli operators are $\sigma^i_p$, defined via
\begin{equation}
S^{i}_p = \frac{1}{2} \, \sigma^{i}_p\,.
\end{equation}
Paulis at different sites commute, while on-site they satisfy
\begin{equation}
\sigma^{i}_p \, \sigma^{j}_{p} = \delta_{ij} \, 1 + \mathrm{i} \, \epsilon_{ijk} \, \sigma^{k}_p.
\end{equation}
The Majorana and Pauli representations are especially convenient for the present work because both sets of operators are Hermitian involutions. This makes it natural to formulate the JW and parity transformations directly at the operator level and to compose them into an explicit SFS mapping. Accordingly, we will express the original spin Hamiltonian in Pauli form, map it to Majorana operators using JW transformation, and then apply the parity transformation to obtain the transformed spin Hamiltonian.

\subsection{Jordan-Wigner Transformation}
Motivated by the correspondence between spin and fermionic operator structures discussed above, we now introduce the JW transformation, first proposed by Jordan and Wigner in 1928 \cite{JordanWigner1928}, which provides an explicit map between spinless fermionic operators to spin operators, which we write as
\begin{subequations}
\begin{align}
c_p^\dagger &\JWmaps S_p^+ \, \tilde{\phi}_p,
\\
c_p &\JWmaps S_p^- \, \tilde{\phi}_p,
\\
\tilde{\phi}_p &= \tilde{\phi}_p^{\dagger}=\prod_{q<p} (-2 \, S^z_q)
\end{align}
\end{subequations}
where $\tilde{\phi}_p$ is the JW string, a many-body operator responsible for yielding the correct algebra: where $S_p^+$ and $S_q^-$ commute for $p \ne q$, $S_p^+ \, \tilde{\phi}_p$ and $S_q^- \, \tilde{\phi}_q$ anticommute. The above mapping suggests that we map the states as
\begin{subequations}
\begin{align}
\ket{0_p} &\JWmaps \ket{\downarrow_p},
\\
\ket{1_p} &\JWmaps \ket{\uparrow_p}.
\end{align}
\end{subequations}
We can invert this transformation to map spin operators to fermionic operators
\begin{subequations}
\begin{align}
S_p^+ &\JWmaps c_p^\dagger \, \phi_p,
\\
S_p^- &\JWmaps c_p \, \phi_p,
\\
S_p^z &\JWmaps n_p - \frac{1}{2},
\\
\phi_p &= \phi_p^{\dagger}=\prod_{q<p} e^{\mathrm{i} \, \pi \, n_q}=\prod_{q<p}  (1 - 2\, n_q).
\end{align}
\end{subequations}

The JW transformation provides a particularly natural starting point because it gives an explicit correspondence between spin and fermionic operators while preserving the required algebra through string operators. An important feature of this mapping is that $S_p^z$ is mapped to the fermion number operator $n_p$ up to a constant shift, so that total $S^z$ symmetry in the spin representation becomes particle-number symmetry in the fermionic representation. At the same time, the mapping is not purely local: the transformed operator on site $p$ depends on all sites preceding $p$ in the chosen ordering. Although this ordering has no effect on exact results, it can influence approximate calculations by changing the ordering of the spins or fermions in the transformed Hamiltonian.

In the present work, it is most convenient to write the JW transformation in terms of Majorana and Pauli operators, since this form connects directly to the parity transformation used in the second step of the SFS construction. In this notation, the JW transformation reads
\begin{subequations}
\begin{align}
\gamma_{1,p} &\JWmaps \sigma^x_p \, \tilde{\phi}_p,
\\
\gamma_{2,p} &\JWmaps \sigma^y_p \, \tilde{\phi}_p,
\\
\tilde{\phi}_p &= \prod_{q<p} \left(-\sigma^z_q\right),
\end{align}
\end{subequations}
while the inverse transformation is
\begin{subequations}
\label{Eqn:JWSpinToFermion}
\begin{align}
\sigma^x_p &\JWmaps \gamma_{1,p} \,\phi_p,
\\
\sigma^y_p &\JWmaps \gamma_{2,p} \,\phi_p,
\\
\sigma^z_p &\JWmaps -\mathrm{i} \, \gamma_{1,p} \, \gamma_{2,p},
\\
\phi_p &= \prod_{q<p} (\mathrm{i} \, \gamma_{1,q} \, \gamma_{2,q}).
\end{align}
\end{subequations}

\subsection{Parity Transformation}
The parity transformation\cite{Seeley2012BK} may be viewed as complementary to the JW transformation. In the Jordan--Wigner mapping, the local occupation of a fermionic mode is encoded directly on the corresponding spin site, while the fermionic sign information is carried by a nonlocal string. The parity transformation distributes this information in a different way: parity is encoded locally, while occupation is represented nonlocally. This makes it a natural second step in the present SFS construction. We have the parity transformation
\begin{subequations}
\label{Eqn:ParityFermionToSpin}
\begin{align}
\gamma_{1,p} &\Paritymaps \psi_p \, \sigma^z_{p-1},
\\
\gamma_{2,p} &\Paritymaps -\mathrm{i} \, \psi_p\, \sigma^z_p,
\\
\psi_p &= \prod_{q \ge p} \sigma^x_q.
\end{align}
\end{subequations}
The inverse transformation is
\begin{subequations}
\begin{align}
\sigma_p^x &\Paritymaps \mathrm{i} \, \gamma_{2,p} \, \gamma_{1,p+1},
\\
\sigma_p^y &\Paritymaps \mathrm{i} \, \phi_p \, \gamma_{1,p} \, \gamma_{1,p+1},
\\
\sigma_p^z &\Paritymaps \mathrm{i} \, \phi_p \, \gamma_{1,p} \, \gamma_{2,p}.
\end{align}
\end{subequations}
For a system with $M$ sites, we define
\begin{align}
\sigma_0^z &\equiv 1,
&
\gamma_{1,M+1} &\equiv \hat{\Pi} = e^{i\pi \hat{N}},
\end{align}
where $\hat{\Pi}$ is the global fermion-number parity operator, which acts on the eigenstates of the number operator with eigenvalue $+1$ if the state has an even number of particles and $-1$ for an odd number of particles. We use the symbol $\Pi$ to refer to the parity transformation. Unlike the JW transformation, which treats the $z$ axis as special because the $\sigma^z_p$ operator does not get a string when mapped, the parity transformation singles out the $x$ axis. This difference is purely conventional; it is convenient to introduce a rotated parity transformation, denoted by $\Pi_R$, that instead singles out the $z$-axis. This rotated form is obtained by the cyclic relabeling $\{\sigma_p^x,\sigma_p^y,\sigma_p^z\} \to \{\sigma_p^z,\sigma_p^x,\sigma_p^y\}$, so that 

\begin{subequations}
\begin{align}
\gamma_{1,p} &\RotatedParitymaps  \ \, \psi_p^z \, \sigma_{p-1}^y,
\\
\gamma_{2,p} &\RotatedParitymaps -\mathrm{i} \, \psi_p^z \, \sigma_p^y,
\\
\psi_p^z &= \prod_{q\ge p}\sigma_q^z,
\\
\sigma_p^x  &\RotatedParitymaps \mathrm{i} \, \phi_p \, \gamma_{1,p} \, \gamma_{1,p+1},
\\
\sigma_p^y  &\RotatedParitymaps \mathrm{i} \, \phi_p \, \gamma_{1,p} \, \gamma_{2,p},
\\
\sigma_p^z  &\RotatedParitymaps \mathrm{i} \, \gamma_{2,p} \, \gamma_{1,p+1},
\end{align}
\end{subequations}
with
\begin{equation}
\sigma_0^y \equiv 1.
\end{equation}

In the present work, both $\Pi$ and $\Pi_R$ are useful. However, $\Pi_R$ is particularly natural when composing with the JW transformation, since both mappings then distinguish the $z$-axis.

\section{Spin-Fermion-Spin Transformations}
The key idea in this manuscript is to compose a spin to fermion transformation with another transformation from fermions back to spins. If each transformation is faithful, the result becomes a canonical transformation from spin 1/2 to spin 1/2.  Since we are interested in spin Hamiltonians, we first determine how the local Pauli operators transform under the SFS mapping. We then illustrate these relations using the 1D XXZ Hamiltonian. 

As an example, let us convert Pauli operators to Majoranas using the JW transformation of Eq.\ref{Eqn:JWSpinToFermion}, and then convert those Majoranas back to Paulis using the $\Pi$ transformation of Eq. \ref{Eqn:ParityFermionToSpin}. 

The JW string contains products like $\mathrm{i} \, \gamma_{1,p} \, \gamma_{2,p}$ so it makes sense to start there.  We have
\begin{equation}
\mathrm{i} \, \gamma_{1,p} \, \gamma_{2,p} \Paritymaps \mathrm{i} \, \psi_p \, \sigma^z_{p-1} \, \left(-\mathrm{i}\right) \, \psi_p \, \sigma^z_{p}.
\end{equation}
Now, note that $\psi_p$ runs over sites $q \ge p$ and thus does not act on site $p-1$. Since Pauli operators on different sites commute, $\psi_p$ commutes with $\sigma^z_{p-1}$, so
\begin{equation}
\mathrm{i} \, \gamma_{1,p} \, \gamma_{2,p} \Paritymaps \psi_p^2 \, \sigma^z_{p-1} \, \sigma^z_{p}.
\end{equation}
Because the individual Pauli operators are commuting involutions ($\sigma_p^x \, \sigma_p^x = 1$) the entire string satisfies $\psi_p^2 = 1$, so we have simply
\begin{equation}
\mathrm{i} \, \gamma_{1,p} \, \gamma_{2,p} \Paritymaps \sigma^z_{p-1} \, \sigma^z_{p}.
\end{equation}

With this in hand, we can check the mapping of the JW string:
\begin{equation}
\phi_p = \prod_{q<p} \left(\mathrm{i} \, \gamma_{1,q} \, \gamma_{2,q}\right) \Paritymaps \prod_{q<p} \left(\sigma^z_{q-1} \, \sigma^z_q\right).
\end{equation}
One can easily verify that
\begin{equation}
\phi_p \Paritymaps \sigma_{p-1}^z.
\end{equation}
 Therefore we have
\begin{subequations}
\begin{align}
\sigma_p^x &\JWmaps \gamma_{1,p} \, \phi_p \Paritymaps \left(\psi_p \, \sigma_{p-1}^z\right) \, \left(\sigma_{p-1}^z\right),
\\
\sigma_p^y &\JWmaps \gamma_{2,p} \, \phi_p \Paritymaps \left(-\mathrm{i} \, \psi_p \, \sigma_p^z\right) \, \left(\sigma_{p-1}^z\right),
\\
\sigma_p^z &\JWmaps -\mathrm{i} \, \gamma_{1,p} \, \gamma_{2,p} \Paritymaps -\mathrm{i} \, \left(\psi_p \, \sigma_{p-1}^z\right) \, \left(-\mathrm{i} \, \psi_p \, \sigma_p^z\right).
\end{align}
\end{subequations}
Simplifying, we obtain the result of the composite mapping:
\begin{subequations}
\begin{align}
\sigma_p^x &\JWParitymaps \tilde{\sigma}_p^x = \psi_p,
\\
\sigma_p^y &\JWParitymaps \tilde{\sigma}_p^y = -\mathrm{i} \, \psi_p \, \sigma_{p-1}^z \, \sigma_p^z,
\\
\sigma_p^z &\JWParitymaps \tilde{\sigma}_p^z = -\sigma_{p-1}^z \, \sigma_p^z.
\end{align}
\end{subequations}
Here, the notation $\Pi \circ \mathrm{JW}$ denotes the composite mapping, where the JW transformation is applied first and is followed by the $\Pi$ transformation. Recall that we have defined $\psi_p$ in Eq. \ref{Eqn:ParityFermionToSpin} and that we use the notational shorthand $\sigma_0^z = 1$.

The composite $\Pi\circ\mathrm{JW}$ mapping has an explicit
representation as a Clifford\cite{Gottesman1997Stabilizer,NielsenChuang2010QuantumComputation,Aaronson2004ImprovedStabilizer} circuit connecting consecutive sites in a chosen ordering. We define
\begin{equation}
C_p
=
\operatorname{CNOT}_{p\rightarrow p+1},
\end{equation}

where $\operatorname{CNOT}_{p\rightarrow p+1}$ denotes a
controlled-NOT\cite{NielsenChuang2010QuantumComputation} gate with site $p$ as the control and site $p+1$
as the target; on computational-basis states it acts as
$|a,b\rangle\mapsto|a,a\oplus b\rangle$, where $\oplus$ denotes
addition modulo two.

For a system of $M$ spin sites, we define the ordered CNOT staircase 
\begin{align}
U_{\mathrm C}
=
C_{M-1}C_{M-2}\cdots C_1
=
\operatorname{CNOT}_{M-1\rightarrow M}
\cdots
\operatorname{CNOT}_{1\rightarrow2},
\label{Eqn:CNOTStaircase}
\end{align}
where the rightmost gate acts first. We refer to $U_{\mathrm C}$ as a CNOT staircase because the
control and target are shifted successively along the chosen site
ordering,
\[
1\rightarrow2,\quad
2\rightarrow3,\quad\ldots,\quad
M-1\rightarrow M.
\]  
We also define the layer
of local spin flips
\begin{equation}
U_X=\prod_{p=1}^{M}\sigma_p^x.
\end{equation}
The unitary implementing the $\Pi\circ\mathrm{JW}$
transformation is
\begin{equation}
U_{\Pi\circ\mathrm{JW}}
= U_{\mathrm C}U_X.
\label{Eqn:SFSUnitary}
\end{equation}
It satisfies
\begin{subequations}
\begin{align}
U_{\Pi\circ\mathrm{JW}}\,
\sigma_p^x\,
U_{\Pi\circ\mathrm{JW}}^\dagger
&=
\prod_{q=p}^{M}\sigma_q^x
=
\psi_p,
\\
U_{\Pi\circ\mathrm{JW}}\,
\sigma_p^z\,
U_{\Pi\circ\mathrm{JW}}^\dagger
&=
-\sigma_{p-1}^z\sigma_p^z,
\end{align}
\end{subequations}
while the transformation of $\sigma_p^y$ follows from
$\sigma_p^y=\mathrm{i}\sigma_p^x\sigma_p^z$.  A gate-by-gate
derivation is given in Appendix~\ref{App:SFSCircuit}.
It is a Clifford circuit since the circuit maps Pauli operators to Pauli products under
conjugation.

This circuit also provides a simple interpretation of the
transformed operators. For $p>1$, the transformed operator
$-\sigma_{p-1}^z\sigma_p^z$ distinguishes aligned and
antialigned spins on consecutive sites in the chosen ordering and thus has a domain-wall-like interpretation, up to the chosen sign convention. For
$p=1$, the convention $\sigma_0^z=1$ instead defines a boundary
variable. In contrast, the conjugate operator $\sigma_p^x$
becomes the extended string
$\prod_{q=p}^{M}\sigma_q^x$. 
The $\Pi\circ\mathrm{JW}$ transformation can therefore be viewed as a domain-wall-type duality implemented by a low-complexity Clifford circuit.

To see how we might use this in practice, consider the XXZ Hamiltonian in 1D. In terms of spins, we may write this Hamiltonian as
\begin{equation}
H_{\mathrm{XXZ}} = \sum_p \left(S_p^x \, S_{{p+1}}^x + S_p^y \, S_{p+1}^y + \Delta \, S_p^z \, S_{p+1}^z\right)
\label{Eqn:XXZHamiltonianOriginal}
\end{equation}
where for open boundary conditions with a chain of length $M$, the sum on $p$ runs from 1 to $M-1$.  Rewriting the Hamiltonian in terms of Pauli operators gives us
\begin{equation}
H_{\mathrm{XXZ}} = \frac{1}{4} \, \sum_p \left(\sigma_p^x \, \sigma_{{p+1}}^x + \sigma_p^y \, \sigma_{p+1}^y + \Delta \, \sigma_p^z \, \sigma_{p+1}^z\right).
\end{equation}
After the sequential spin-fermion-spin transformation, this becomes
\begin{align}
H_{\mathrm{XXZ}} \JWParitymaps \frac{1}{4} \, \sum_{p=1}^{M-1} \Big(&\psi_p \, \psi_{p+1} - \psi_p \, \sigma_{p-1}^z \, \sigma_p^z \, \psi_{p+1} \, \sigma_p^z \, \sigma_{p+1}^z
\nonumber
\\
&+\Delta \, \sigma_{p-1}^z \, \sigma_p^z \, \sigma_p^z \, \sigma_{p+1}^z\Big)
\end{align}
To simplify, we use that $\psi_{p} = \psi_{p+1} \, \sigma_p^x$, that $\psi_{p+1}$ depends only on sites $p+1$ to $M$ and therefore commutes with operators on sites 1 to $p$, and that $\psi_{p+1}^2 = 1$, and so too for operators like $\sigma_p^z$. We arrive at
\begin{align}
H_{\mathrm{XXZ}} \JWParitymaps \frac{1}{4} \, \sum_{p=1}^{M-1} \Big(&\sigma_p^x - \sigma_{p-1}^z \,  \sigma_p^x \, \sigma_{p+1}^z
\nonumber
\\
&+ \Delta \, \sigma_{p-1}^z \, \sigma_{p+1}^z\Big).
\end{align}
If we wish, we can, finally, return to spin operators:
\begin{align}
H_{\mathrm{XXZ}} \JWParitymaps \sum_{p=1}^{M-1} \Big(&\frac{1}{2} \, S_p^x - 2 \, S_{p-1}^z \,  S_p^x \, S_{p+1}^z
\nonumber
\\
&+ \Delta \, S_{p-1}^z \, S_{p+1}^z\Big),
\label{Eqn:XXZHamiltonianMapped}
\end{align}

where at the boundary \begin{equation}
    S^z_{0} \equiv 1/2
\end{equation}

The circuit representation makes the equivalence of the original
and transformed Hamiltonians explicit:
\begin{equation}
H_{\mathrm{XXZ}}^{\mathrm{SFS}}
=
U_{\Pi\circ\mathrm{JW}}\,
H_{\mathrm{XXZ}}\,
U_{\Pi\circ\mathrm{JW}}^\dagger.
\label{Eqn:SFSXXZUnitary}
\end{equation}
The two Hamiltonians are therefore isospectral, with their
eigenstates related by
$|\widetilde{\Psi}_n\rangle
=
U_{\Pi\circ\mathrm{JW}}|\Psi_n\rangle$.
We could use the rotated parity mapping instead, but since this amounts to a global spin rotation of the final transformed Hamiltonian, it is physically inconsequential.  

In what follows, we will consider four different approaches: an initial JW transformation followed by a parity transformation ($\Pi \, \circ \, \mathrm{JW}$), an initial JW transformation followed by a rotated parity transformation ($\Pi_R \, \circ \, \mathrm{JW}$), an initial parity transformation followed by a JW transformation ($\mathrm{JW} \, \circ \, \Pi)$, and an initial rotated parity transformation followed by a JW transformation ($\mathrm{JW} \, \circ \, \Pi_R$). Detailed expressions for the various canonical transformations appear in Appendix~\ref{sfs_expressions}.
Since the transformation is implemented by unitary conjugation, the transformed operators satisfy the same on-site $\mathfrak{su(2)}$ algebra and commute on different sites; a gate-by-gate verification is provided in Appendix~\ref{App:SFSCircuit}.

\section{Spin Mean Field Theory}
The mean-field theory for spin 1/2 is a spin version of the Bardeen-Cooper-Schrieffer (BCS)\cite{Bardeen1957Theory} wavefunction. We write it as
\begin{equation}
\ket{\mathrm{sBCS}} = \prod_p \left(u_p + v_p \, S_p^+\right) \, \ket{\Downarrow}
\end{equation} 
where $u_p$ and $v_p$ are variational parameters with the normalization condition
\begin{equation}
u_p \, u_p^\star + v_p \, v_p^\star = 1
\end{equation}
and where $\ket{\Downarrow}$ is the spin vacuum where each site has $\downarrow$-spin so that
\begin{equation}
S_p^- \ket{\Downarrow} = 0 \quad \forall p.
\end{equation}
The ansatz can be viewed as a product of local spinors, in which each site is described by a superposition of a $\downarrow$-spin state and an $\uparrow$-spin state or, if one prefers, as a product in which each site is $\downarrow$-spin in some direction, not necessarily the same on each site.  In this sense, the sBCS wavefunction plays the role of a mean-field reference for spin Hamiltonians, analogous to the usual BCS wavefunction in fermionic systems. Although the ansatz is separable in the site basis, it captures nontrivial broken-symmetry physics through the variational amplitudes $\{u_p,v_p\}$.

The spin BCS (sBCS) wavefunction is a superposition of spin configurations with different $S_z$ quantum numbers, and thus is not generally an eigenstate of the $S_z$ operator. For Hamiltonians which have $S_z$ symmetry, this means that sBCS allows symmetry-broken solutions. However, this symmetry breaking can be useful as it enlarges the variational manifold and can provide a lower-energy reference state. When needed, broken symmetry can be restored by projection onto a fixed sector $S_z$. In this work, energies reported are the minima in the full spin Hilbert space rather than being restricted to any subspace of a specific $S_z$. 

Since the sBCS state will later serve as a reference for CC theory, it is worth introducing a quasispin basis in which the sBCS wavefunction is a vacuum. The parameters $u_p$ and $v_p$ define the
local quasispin basis transformation\cite{Bardeen1957Theory,Bogoliubov1958Superconductivity,Valatin1958Superconductivity}
\begin{subequations}
\begin{align}
\ket{\downarrow_p'}
&=
u_p\ket{\downarrow_p}
+
v_p\ket{\uparrow_p},
\\
\ket{\uparrow_p'}
&=
-v_p^{*}\ket{\downarrow_p}
+
u_p^{*}\ket{\uparrow_p}.
\end{align}
\end{subequations}
These states are orthonormal provided
$|u_p|^2+|v_p|^2= 1$.
We define the corresponding local quasispin raising and lowering
operators as
\begin{subequations}
\begin{align}
\mathcal{S}_p^{+}
&=
\ket{\uparrow_p'}\bra{\downarrow_p'},
\\
\mathcal{S}_p^{-}
&=
\ket{\downarrow_p'}\bra{\uparrow_p'}.
\end{align}
\end{subequations}

In this basis, the sBCS wavefunction can be
written as
\begin{equation}
\ket{\mathrm{sBCS}}
=
\bigotimes_{p=1}^M \ket{\downarrow_p'}.
\end{equation}
 
The sBCS reference therefore satisfies
\begin{equation}
\mathcal{S}_p^{-}\ket{\mathrm{sBCS}}=0
\qquad \forall p,
\end{equation}
and may be regarded as the vacuum of the quasispin
operators.

The corresponding spin-mean-field energy is obtained as
\begin{equation}
E_{\mathrm{sBCS}} = \frac{\bra{\mathrm{sBCS}} H \ket{\mathrm{sBCS}}}{\braket{\mathrm{sBCS}|\mathrm{sBCS}}}
\end{equation}
which is minimized with respect to the variational parameters subject to the normalization constraints.

In practice, it is convenient to parameterize the amplitudes in terms of angles,
\begin{align}
u_p &= \cos(\theta_p),   &v_p &= \sin(\theta_p)
\end{align}
for the real case, or more generally
\begin{align}
u_p &= \cos(\theta_p) ,  &v_p &= \sin(\theta_p) \, \mathrm{e}^{\mathrm{i} \, \varphi_p}
\end{align}
when complex variational parameters are allowed.

\begin{figure}[t]
\includegraphics[width=0.95\columnwidth]{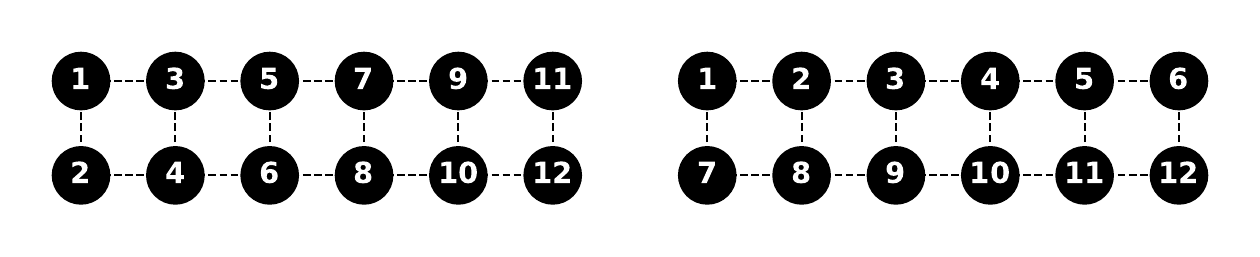}
\caption{Labeling schemes for 2D lattices. Two labeling schemes for the same finite $2\times n$ lattice, Left: the $2\times n$ labeling for $n=6$. Right: the $n\times 2$ labeling for $n=6$.}
\label{fig:labeling_scheme}
\end{figure}

As shown in Appendix~\ref{app:sfs_scaling}, the SFS transformation does not increase the asymptotic cost of an efficiently implemented mean-field energy evaluation. By exploiting the product structure of sBCS and reusing intermediate string products, the cost remains $\mathcal{O}(M)$ for finite-range Hamiltonians and $\mathcal{O}(M^2)$ for generic two-body Hamiltonians, although
the computational prefactor may increase.

\section{Results}
\label{SFS_MF_Results}
We now examine the performance of the SFS representation at the mean-field level. Although an SFS transformation leaves the exact spectrum unchanged, the energy corresponding to approximate wavefunctions is not invariant under these transformations. This is because the transformation changes the operator representation of both the Hamiltonian and its symmetries, while a restricted ansatz need not be closed under the corresponding unitary map.

Let $\mathcal{M}_{\mathrm{sBCS}}$ denote the manifold of sBCS
product states $\ket\Phi$ defined in Sec.~IV. For an SFS transformation
implemented by $U_{\mathrm{SFS}}$, the transformed mean-field
energy may be written as

\begin{align}
E_{\mathrm{SFS\text{-}sBCS}}
&=
\min_{\Phi\in\mathcal{M}_{\mathrm{sBCS}}}
\langle\Phi|
U_{\mathrm{SFS}} H U_{\mathrm{SFS}}^\dagger
|\Phi\rangle
\nonumber\\
&=
\min_{\Phi\in\mathcal{M}_{\mathrm{sBCS}}}
\langle
U_{\mathrm{SFS}}^\dagger\Phi
|
H
|
U_{\mathrm{SFS}}^\dagger\Phi
\rangle .
\label{Eqn:SFSDressedAnsatz}
\end{align}
Thus, an sBCS product state in the transformed representation
corresponds to an entangled, circuit-dressed state
$U_{\mathrm{SFS}}^\dagger|\Phi\rangle$ in the original
representation. The original and the SFS-transformed mean-field
references, therefore, optimize over different restricted
variational manifolds.

Conserved quantities are transformed by the same unitary. If
$[H,Q]=0$, then
\begin{equation}
\left[
U_{\mathrm{SFS}}HU_{\mathrm{SFS}}^\dagger,
U_{\mathrm{SFS}}QU_{\mathrm{SFS}}^\dagger
\right]
=
0.
\end{equation}

For example, the original XXZ Hamiltonian conserves the total magnetization
\begin{equation}
Q = \sum_p S_p^z .
\end{equation}
In the $\Pi \circ \mathrm{JW}$ representation,
\begin{equation}
\sigma_p^z \JWParitymaps -\sigma_{p-1}^z \, \sigma_p^z ,
\end{equation}
and therefore the conserved quantity is mapped to
\begin{equation}
Q_{\mathrm{SFS}} = -2 \, \sum_p S_{p-1}^z \, S_p^z .
\end{equation}
The SFS Hamiltonian conserves the image of the original magnetization symmetry. This mapped symmetry provides a useful diagnostic for understanding how mean-field solutions differ between the original and transformed representations. 

In this work, we consider the $J_1-J_2$ Heisenberg model and the spin-XXZ model in one and two dimensions. It is important to note that the results depend on the chosen labeling scheme. In 1D, we use a sequential labeling scheme, where sites on the lattice are assigned sequential numbers: site 1 is connected to site 2, site 2 to site 3, and so on. Similarly, in 2D we consider two sequential site orderings  (see Fig.\ref{fig:labeling_scheme}). Here, \enquote{2D} refers to the finite $2\times n$ rectangular clusters and no extrapolation to
the thermodynamic limit is implied.  From the circuit perspective, changing the site labeling changes
the sequence of control and target sites in the CNOT staircase.
Different orderings therefore define different dressed
variational manifolds
$U_{\mathcal O}^{\dagger}\mathcal{M}_{\mathrm{sBCS}}$.
Although the corresponding transformed Hamiltonians are
isospectral, their approximate mean-field energies need not
coincide because the sBCS manifold is not invariant under these ordering-dependent circuits. We also present results for different SFS mappings, such as $\mathrm{JW} \, \circ \, \Pi$ and $\mathrm{JW} \, \circ \, \Pi_R$, to understand how the rotated choice of coordinates in the mapping affects the sBCS energies. 

For each value of the Hamiltonian parameters, the SFS-sBCS energy was
optimized independently from 50 randomly generated initial
guesses. Each trial was minimized using the BFGS method
with a convergence threshold of $10^{-9}$, and the
lowest-energy converged solution among the 50 trials was
retained. The random initial spinors were generated subject to
the normalization condition
$|u_p|^2+|v_p|^2=1$.
The energy error $\Delta E$ throughout this manuscript is defined relative to the exact diagonalization\begin{equation}
\Delta E
=
E_{\mathrm{method}}-E_{\mathrm{ED}},
\end{equation}where $E_{\mathrm{ED}}$ 
is the exact ground-state energy.

\subsubsection{The $J_1-J_2$ Hamiltonian}
The $J_1-J_2$ Hamiltonian is 
\begin{equation}
H_{J_1-J_2} = J_1 \, \sum_{\langle p q\rangle} \vec{S}_p \cdot\vec{S}_{q} + J_2 \, \sum_{\langle \langle pq\rangle\rangle} \vec{S}_p \cdot\vec{S}_{q}
\end{equation}
where $\langle p q\rangle$ denotes nearest-neighbor sites and  $\langle \langle pq\rangle\rangle$ second-nearest-neighbor sites. The Hamiltonian preserves $S^2$ symmetry, and so is invariant under any global $\mathfrak{su}(2)$ rotation. The physics depends on the ratio $J_2/J_1$, where we take both $ J_2$ and $ J_1$ to be positive. For a 2D square lattice in the thermodynamic limit, the ground state for  $J_2/J_1 \lesssim 0.45$ is a N\'eel antiferromagnet, and for $J_2/J_1 \gtrsim 0.6 $ is a striped antiferromagnetic phase \cite{Liu2023SpinAGP}. In between, the magnetic structure is in a highly spin-frustrated phase \cite{Darradi2008,Gong2014,Richter2015,Wang2013GaplessSpinLiquidJ1J2,Bishop1998PhaseTransitionsJ1J2}. Throughout this work, we fix $J_1=1$ and vary $J_2$\cite{Jiang2012SpinLiquidJ1J2,Murg2009PEPSFrustrated,Yu2012PlaquetteTensorJ1J2}.

\begin{figure*}[t]
\includegraphics[width=0.48\textwidth]{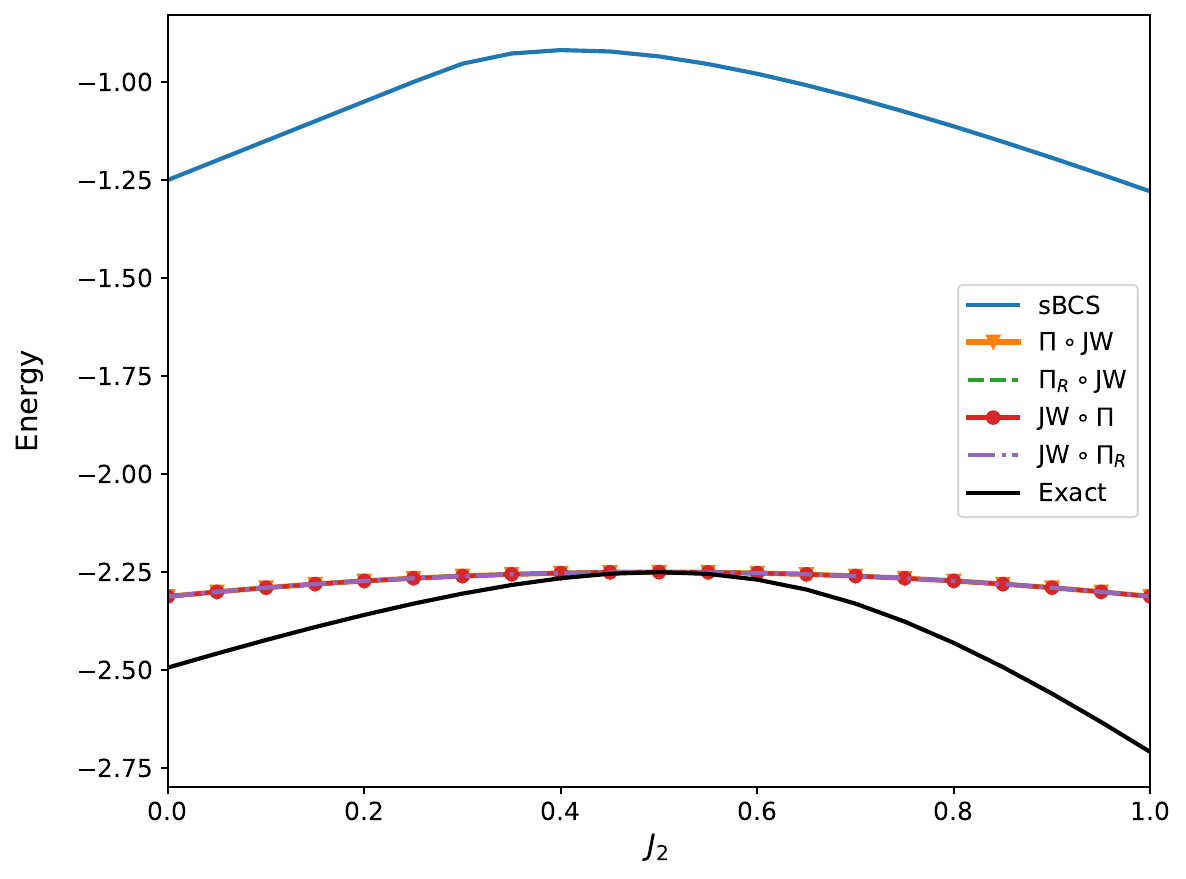}
\hfill
\includegraphics[width=0.48\textwidth]{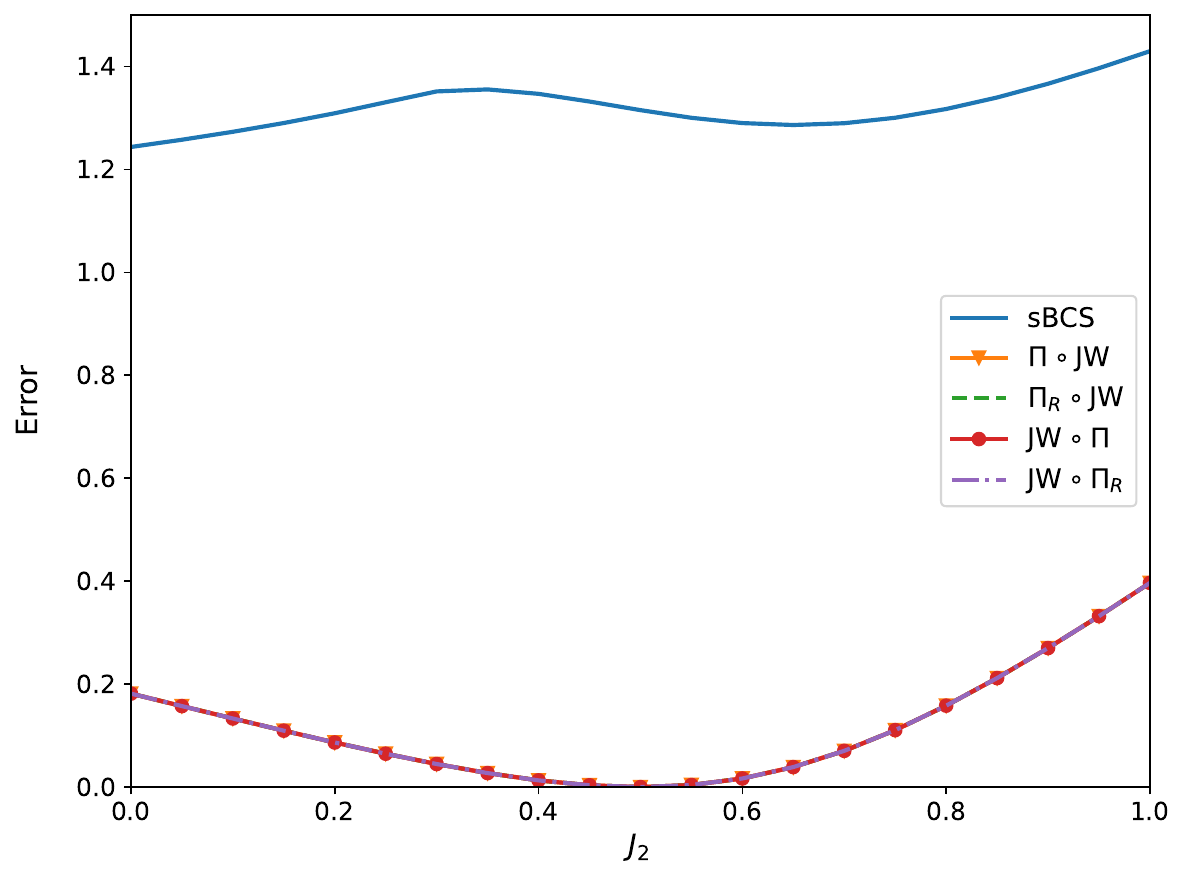}
\caption{Total energies and energy errors for the six-site $J_1$--$J_2$ Hamiltonian with open boundary conditions. Left panel: total energies. Right panel: energy errors relative to the exact result.}
\label{fig:j1j2_6_obc}
\end{figure*}

\begin{figure*}
\includegraphics[width=0.48\textwidth]{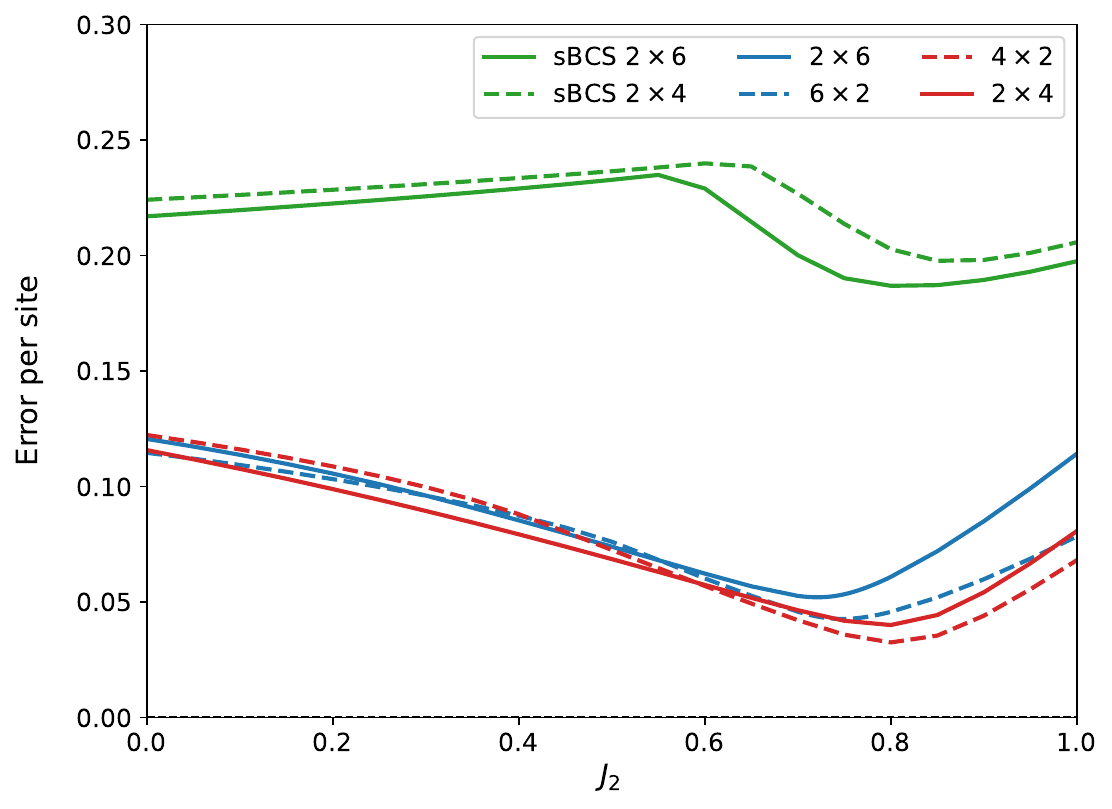}
\hfill
\includegraphics[width=0.48\textwidth]{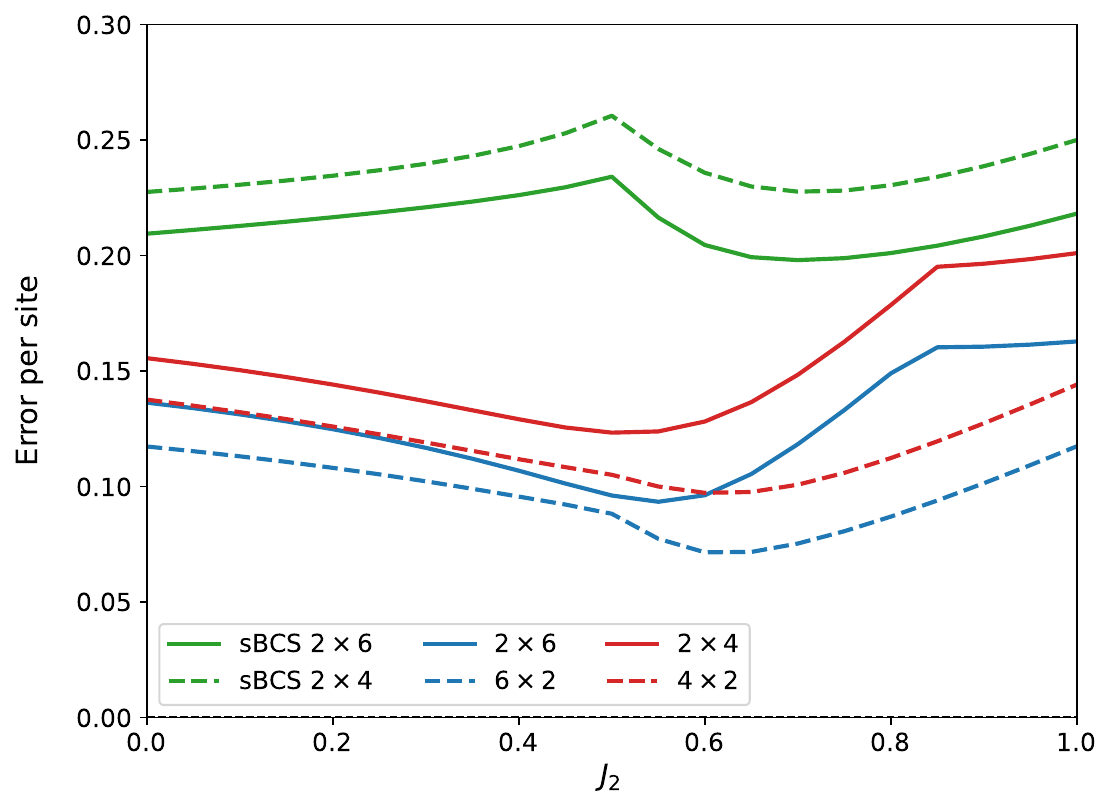}
\caption{Energy errors per site for the $n \times 2$ and $2\times n$ SFS-transformed $J_1$--$J_2$ Hamiltonian relative to exact diagonalization. All SFS curves shown correspond to the
\(\Pi\circ\mathrm{JW}\) transformation; the other SFS variants yield
identical sBCS energies for this isotropic Hamiltonian. Left panel: open boundary conditions. Right panel: periodic boundary conditions.}

\label{fig:j1j2_nx2_obc_pbc}
\end{figure*}

Figure \ref{fig:j1j2_6_obc} shows the total energies and energy errors for the 6-site SFS-transformed $J_1-J_2$ Hamiltonian with open boundary conditions (OBC). We can immediately see that all the SFS-transformed sBCS energies are identical.  That it makes no difference whether we use $\Pi$ or $\Pi_R$ follows from the global $\mathfrak{su}(2)$ symmetry; since the interaction is isotropic, a global rotation of the spin coordinate axes leaves the Hamiltonian unchanged. That $\Pi \, \circ \, \mathrm{JW}$ and $\mathrm{JW} \, \circ \, \Pi$ are identical is less obvious, but in some sense these transformations are duals of one another and we have observed previously that in spin to fermion mappings, they provide identical results for isotropic spin Hamiltonians\cite{Henderson2024FermionicMeanField}.

At $J_2 = 0.5$, the SFS-sBCS energies are exact. This point is the Majumdar--Ghosh point, at which the exact ground state is a dimerized singlet \cite{Majumdar1969}. 

In Fig. \ref{fig:j1j2_nx2_obc_pbc}, we compare results for the $2\times4$ and $2 \times 6$ SFS-transformed $J_1-J_2$ Hamiltonian with both open boundary conditions (OBC) and periodic boundary conditions (PBC). All SFS curves shown correspond to the
\(\Pi\circ\mathrm{JW}\) transformation. The SFS results in 2D are less accurate than in 1D, presumably because strings now contribute. We note that we obtain superior results with OBC than with PBC. 

Because the JW mapping is quasi-one-dimensional (i.e., it is not invariant under relabeling of sites), as is the parity transformation, our results after the SFS transformation depend on the labeling scheme.  We also note that for PBC, this labeling dependence is exacerbated compared to OBC.

\begin{figure*}[t]
    \centering

    \begin{subfigure}[t]{0.48\textwidth}
        \centering
        \includegraphics[width=\linewidth]{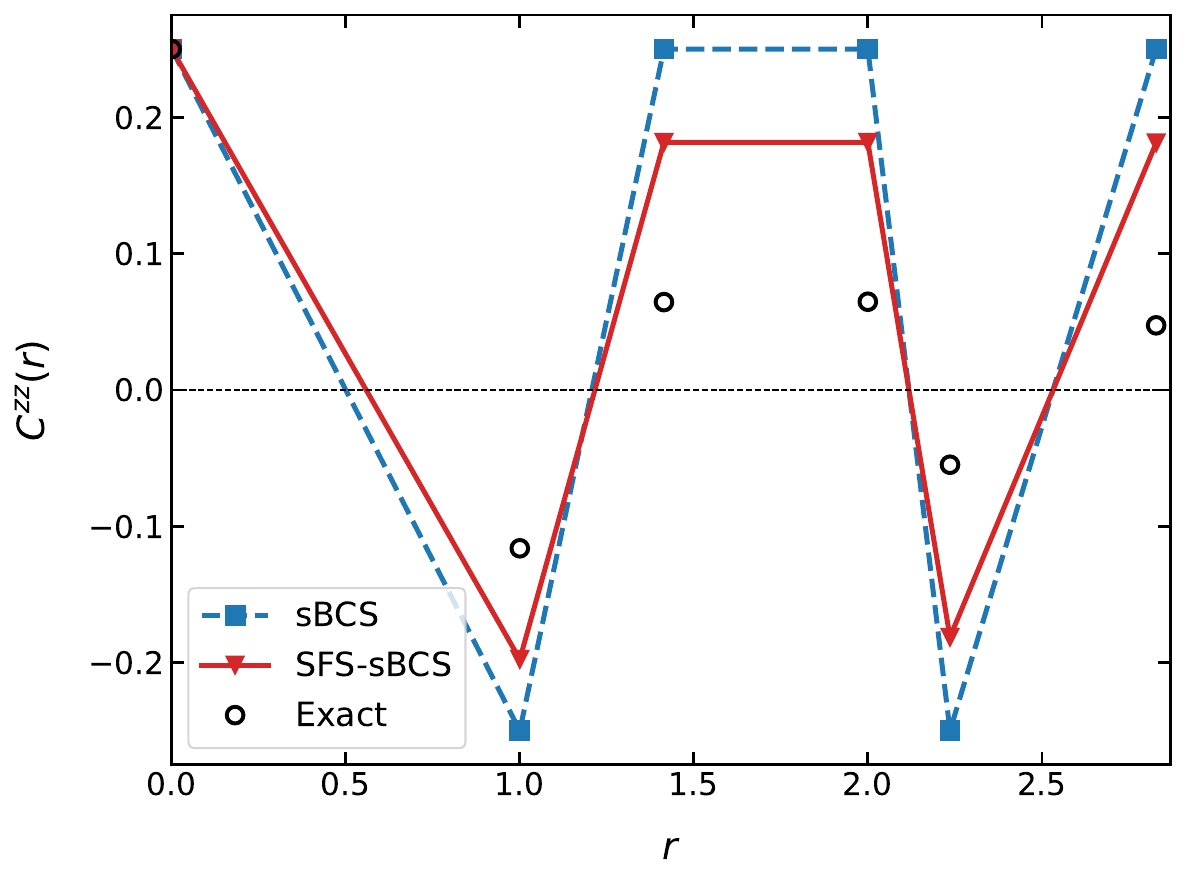}
        \caption{}
        \label{fig:SFS_corr_J2_02}
    \end{subfigure}%
    \hfill
    \begin{subfigure}[t]{0.48\textwidth}
        \centering
        \includegraphics[width=\linewidth]{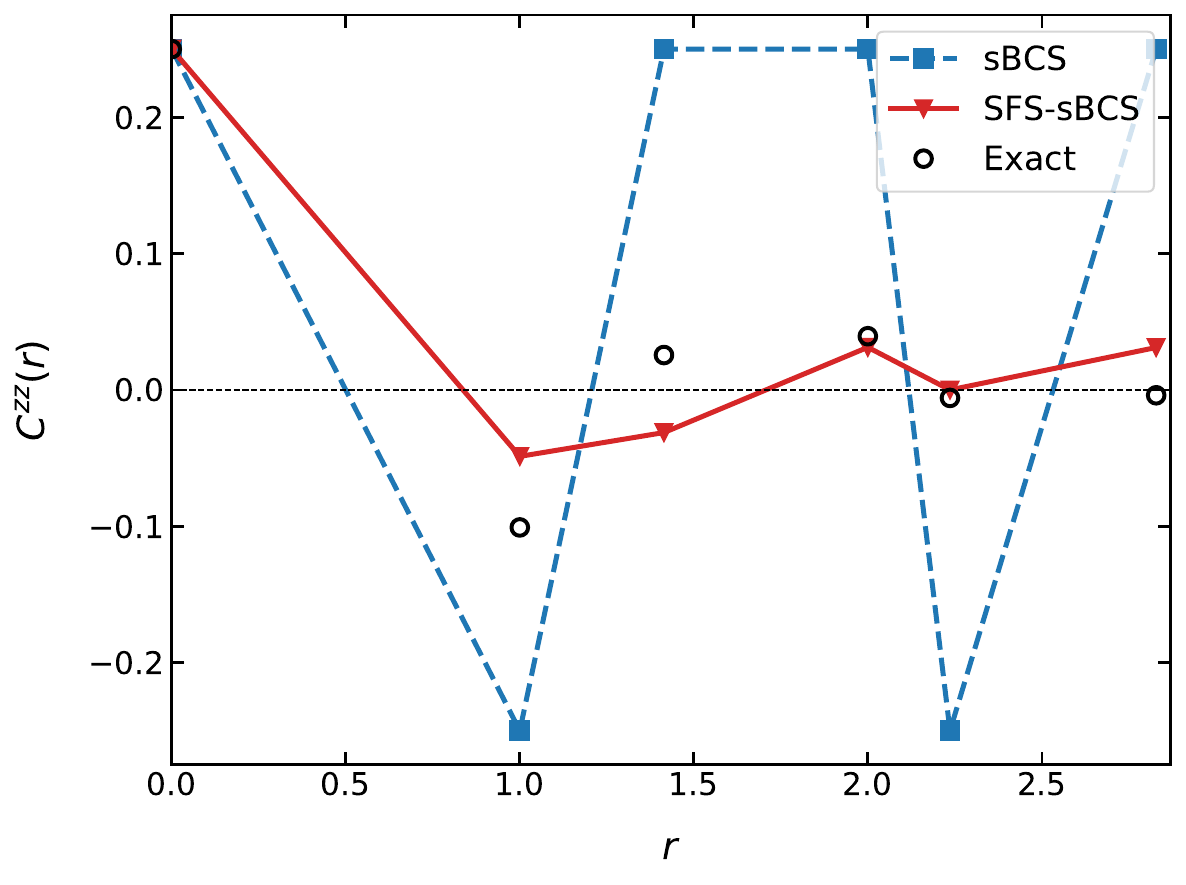}
        \caption{}
        \label{fig:SFS_corr_J2_05}
    \end{subfigure}

    \par\vspace{0.5em}

    \begin{subfigure}[t]{0.48\textwidth}
        \centering
        \includegraphics[width=\linewidth]{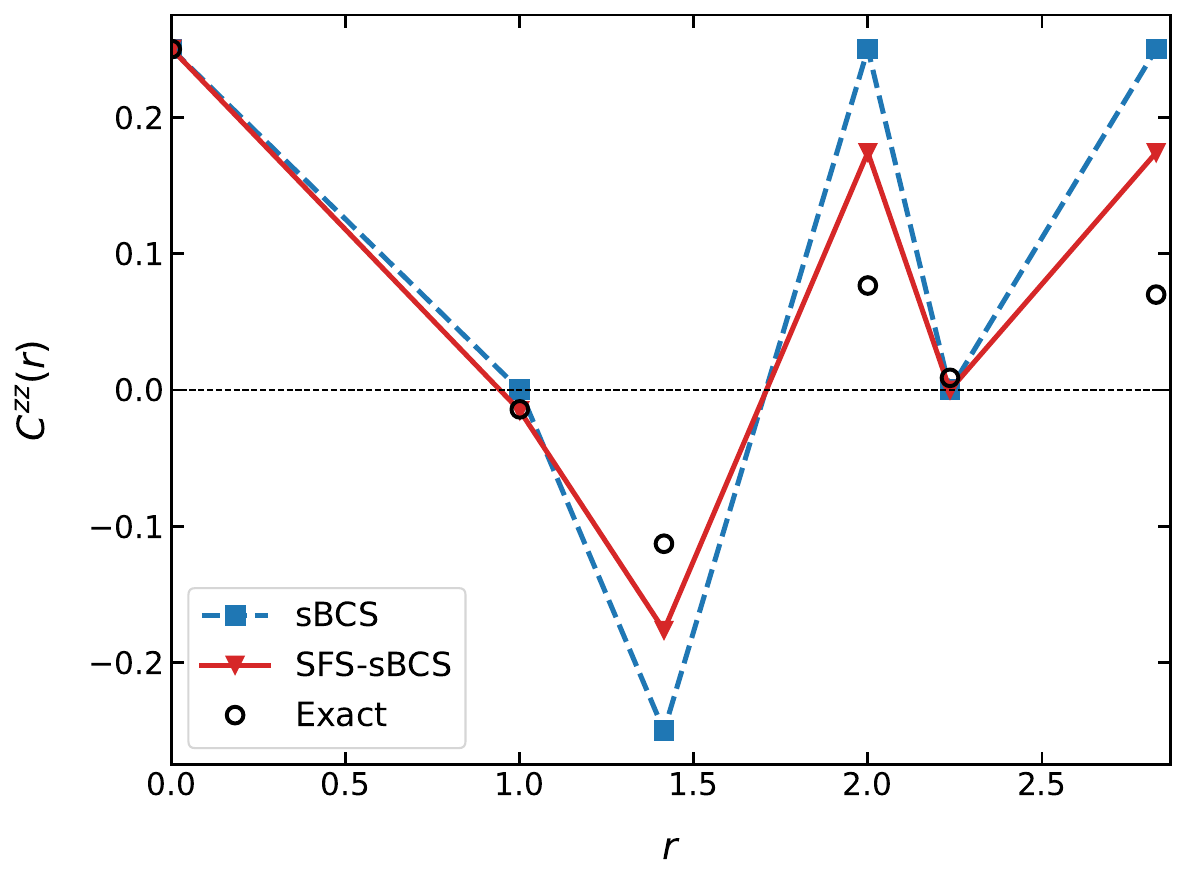}
        \caption{}
        \label{fig:SFS_corr_J2_08}
    \end{subfigure}%
    \hfill
    \begin{subfigure}[t]{0.48\textwidth}
        \centering
        \includegraphics[width=\linewidth]{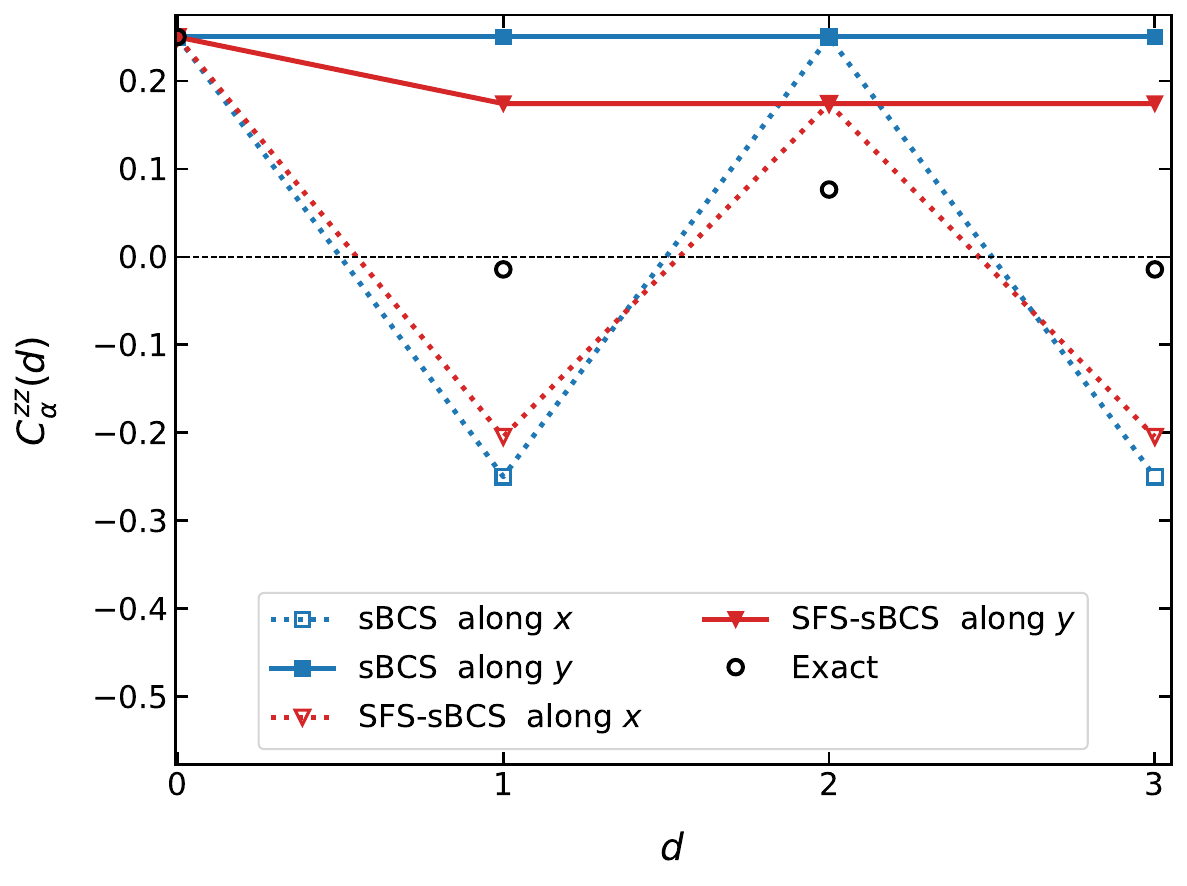}
        \caption{}
        \label{fig:SFS_corr_stripe}
    \end{subfigure}

    \caption{Spin--spin correlation functions for the
    $4\times4$ square-lattice $J_1$--$J_2$ model with
    periodic boundary conditions. Panels (a)--(c) show
    the radial correlations at $J_2=0.2$, $0.5$, and $0.8$,
    respectively. Panel (d) shows the direction-resolved
    correlations in the stripe regime at $J_2=0.8$. SFS mean field corresponds to $\Pi\circ \text{JW}$ transformation.
    }
    \label{fig:SFS_spin_correlations}

\end{figure*}

To examine the  size extensivity of the SFS-sBCS, we also studied the two-dimensional
$J_1$--$J_2$ model on various system sizes, and the results are reported in the supplementary material. The SFS-sBCS energy per site exhibits smooth finite-size behavior, and its improvement over conventional sBCS persists with increasing system
size. These results suggest that the SFS mean-field advantage is not restricted to small systems.

Figure~\ref{fig:SFS_spin_correlations} compares the longitudinal
spin--spin correlation functions $C^{zz}$ obtained with sBCS,
SFS-sBCS, and exact diagonalization. The radial correlation
function is defined as

\begin{equation}
C^{zz}(r)
= \frac{1}{N_r}
\sum_{\substack{p,q\\d_{\mathrm{PBC}}(p,q)=r}}
\left\langle S_p^z S_q^z \right\rangle,
\end{equation}

where $d_{\mathrm{PBC}}(p,q)$ is the minimum-image distance
between sites $p$ and $q$, and $N_r$ is the number of ordered site pairs in the corresponding distance shell.

For SFS-sBCS, the physical longitudinal correlation function
is evaluated using the transformed spin operators,
\begin{equation}
C_{zz}^{\mathrm{SFS}}(p,q)
=
\langle\Phi_{\mathrm{SFS}}|
\widetilde S_p^z\widetilde S_q^z
|\Phi_{\mathrm{SFS}}\rangle,
\end{equation}
where
$\widetilde S_p^z
=U_{\mathrm{SFS}}S_p^zU_{\mathrm{SFS}}^\dagger$
and $|\Phi_{\mathrm{SFS}}\rangle$ is the optimized
sBCS state of the transformed Hamiltonian. From the Fig.~\ref{fig:SFS_spin_correlations}, at $J_2=0.2$, SFS-sBCS preserves the N\'eel correlation pattern and improves the correlation over standard sBCS.
Near the frustrated regime ($J_2=0.5$), the SFS transformation
substantially suppresses the classical correlations, bringing
them closer to the exact results, although discrepancies remain. At $J_2=0.8$, both of the methods resemble a striped phase. To show the striped phase explicitly, we have added a direction-resolved correlations plot (see
Fig.~\ref{fig:SFS_corr_stripe}) which shows that SFS-sBCS reduces the
magnitude of the stripe correlations but retains the anisotropy
absent in the exact finite-cluster ground state.


\subsubsection{The XXZ Hamiltonian}

\begin{figure*}[t]
\includegraphics[width=0.48\textwidth]{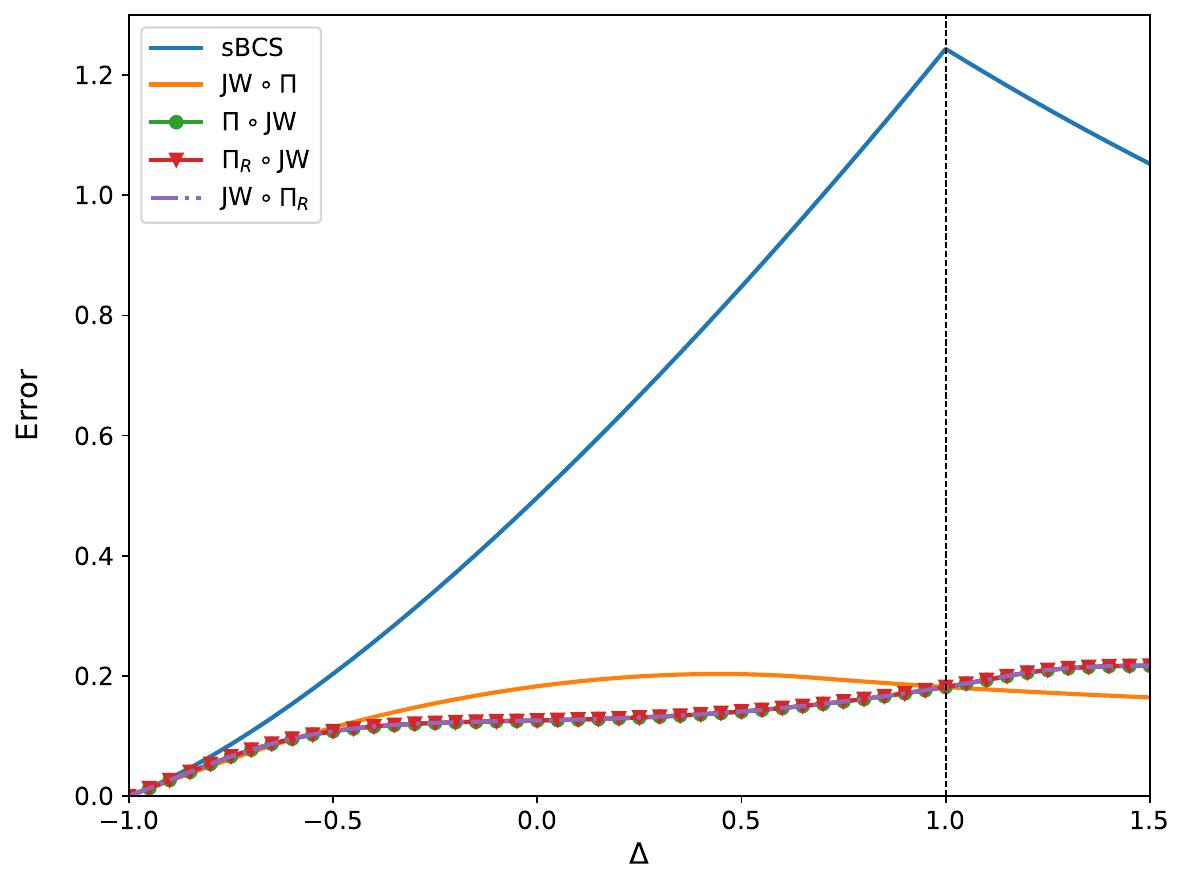}
\hfill
\includegraphics[width=0.48\textwidth]{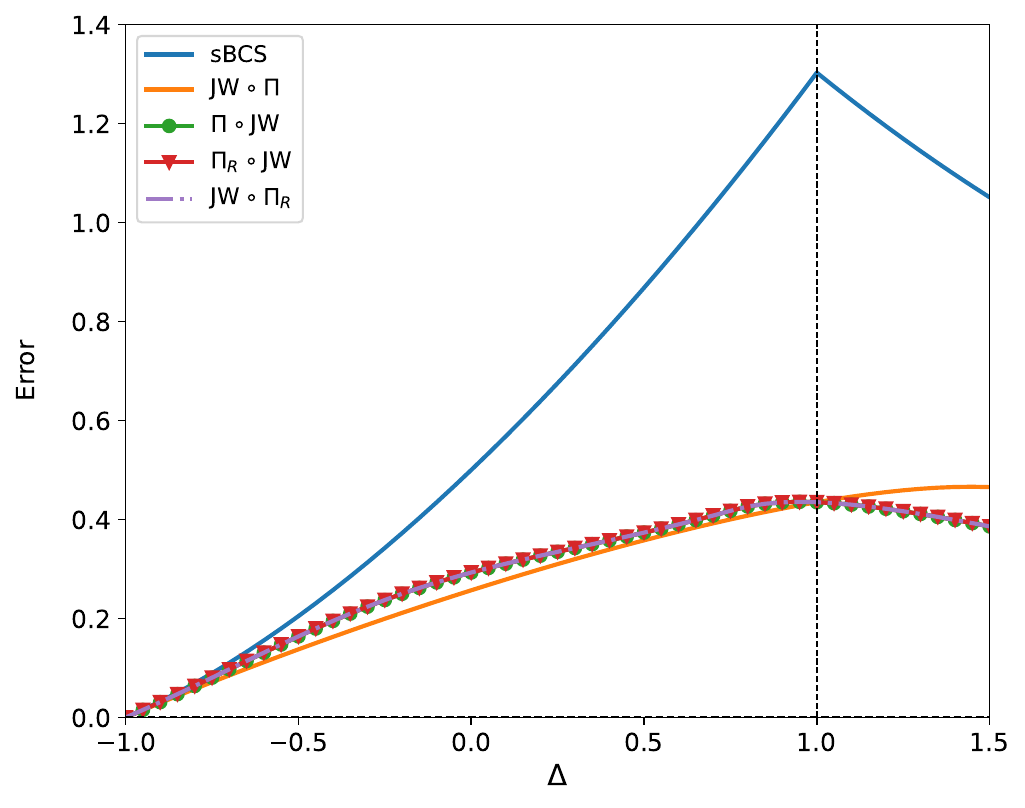}
\caption{Energy errors for the 6-site SFS-transformed XXZ Hamiltonian relative to exact diagonalization. Left panel: open boundary conditions. Right panel: periodic boundary conditions.}
\label{fig:1dxxz}
\end{figure*}

The general XXZ Hamiltonian is given by
\begin{equation}
H_{\mathrm{XXZ}} =\sum_{\langle pq\rangle} \left(S^x_{p} \, S^{x}_{q} + S^y_p \, S^y_q + \Delta \, S^{z}_{p} \, S^{z}_{q} \right).
\end{equation}
The physics of the XXZ model depends on the parameter $\Delta$. The point at $\Delta=1$ is the Heisenberg point at which the Hamiltonian has both $S^2$ and $S_z$ symmetries, but it only conserves $S_z$ for all other $\Delta$. In one dimension, the model is integrable and admits an exact Bethe-ansatz solution, with the isotropic Heisenberg chain solved originally by Bethe and the anisotropic XXZ chain analyzed by Yang ~\cite{Bethe1931BetheAnsatz,Yang1966XXZ2,Yang1966XXZ3}. There exist three phases for this model in the thermodynamic limit\cite{Yang1966XXZ2,Yang1966XXZ3,Giamarchi2004OneDimension}. For $\Delta < -1$, the ground state is in a ferromagnetic order, which occurs for maximal $|S_z|$. In the region $-1 < \Delta < 1$, the ground state is a gapless XY phase, where the spins are in the $xy$ plane \cite{Farnell2004CCMQuantumMagnetism}, whereas for $\Delta > 1$, the ground state is antiferromagnetic with the magnetization along the $z$ axis. The ground state over all $S_z$ sectors occurs at $S_z=0$ for $\Delta>-1$, while different $S_z$ sectors are degenerate with one another at $\Delta =-1$ \cite{Yang1966XXZ3,Giamarchi2004OneDimension}.

First, we consider the 1D case depicted in Fig. \ref{fig:1dxxz} in which the strings drop out and the SFS-transformed Hamiltonian remains local. Note that in PBC, there is a string connecting the first and last sites.

We again see that the sBCS mean-field on the SFS-transformed Hamiltonian yields significantly improved energies compared to the standard sBCS mean-field.  We also observe something new compared to the $J_1$-$J_2$ model: following the parity transformation from spins to fermions with the JW transformation from fermions back to spins ($\mathrm{JW} \, \circ \, \Pi$) gives different results than do the other transformations.  As we have noted, all four transformations are canonical and all yield the same result when the Hamiltonians are solved exactly, but there is no reason they must yield the same result when the Hamiltonians are solved approximately.  Global rotations \textit{after} the SFS transformation leave the final result of the spin mean-field invariant, but global rotations \textit{before} the transformation do not. Again, this is simply because global rotations after the transformation can be compensated for by a general complex spin BCS solution.  The exception is at $\Delta = 1$ where the original Hamiltonian is isotropic and so invariant to global spin rotations.



\begin{figure*}[t]
\includegraphics[width=0.48\textwidth]{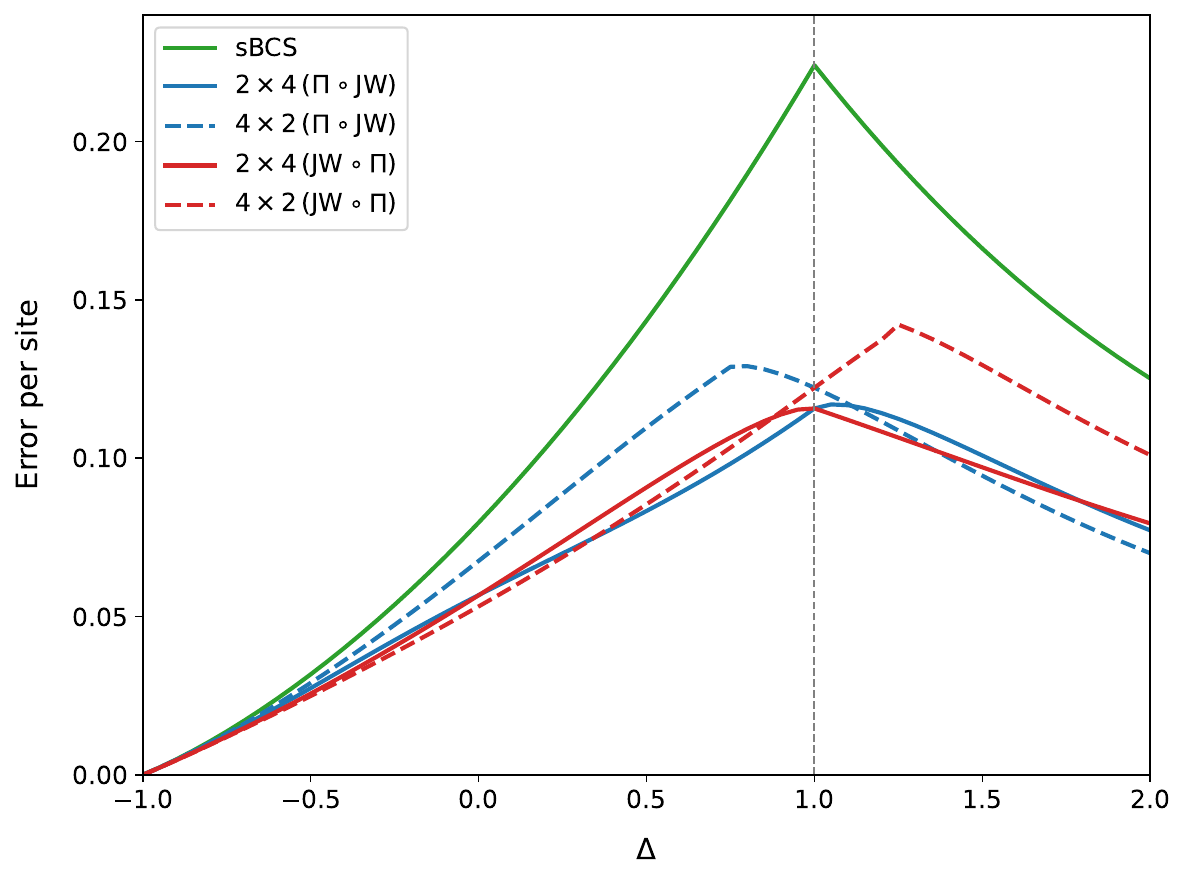}
\hfill
\includegraphics[width=0.48\textwidth]{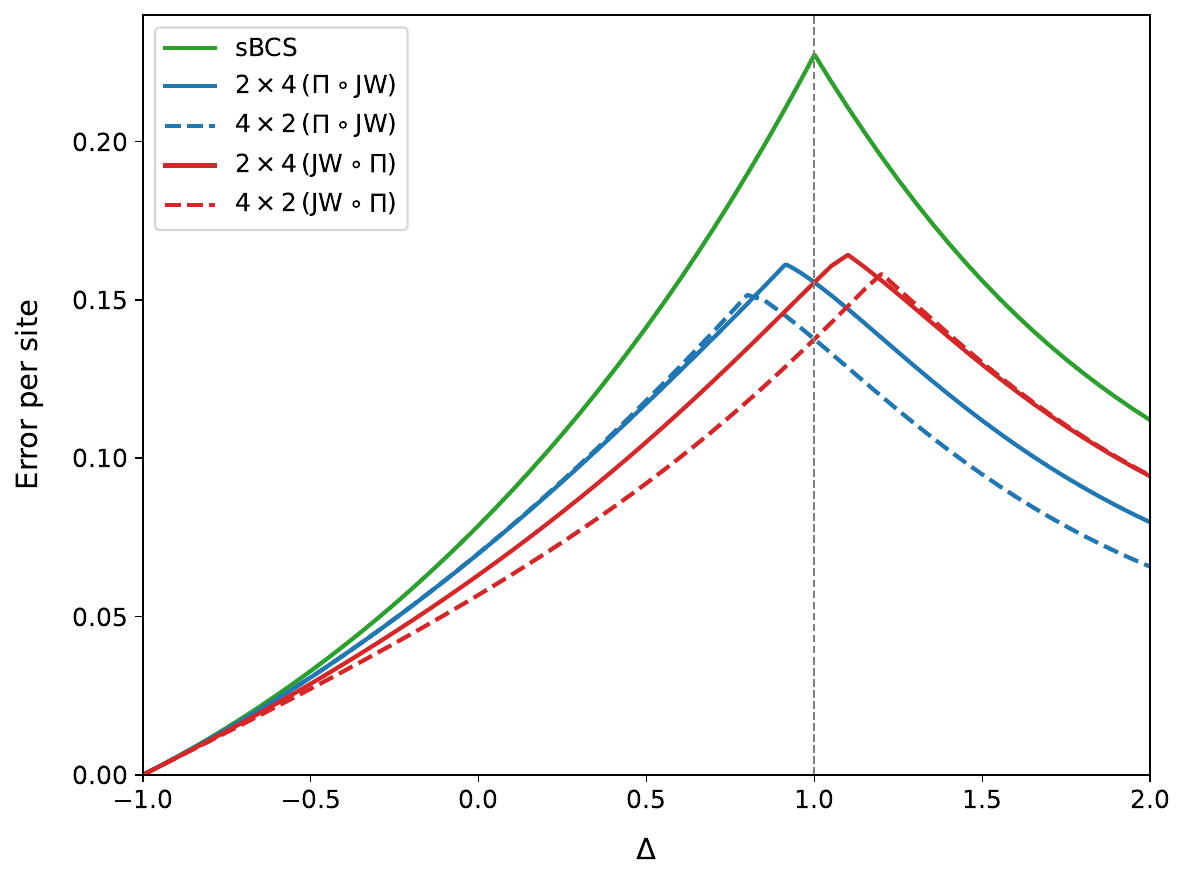}
\caption{Energy errors per site for the $4 \times2 $ and $2\times4$ SFS-transformed XXZ Hamiltonian relative to exact diagonalization. Left panel: Open boundary conditions. Right panel: periodic boundary conditions.}
\label{fig:xxz_2x4_pbc_obc_errors}
\end{figure*}

At $\Delta =-1$, all methods are exact, since sBCS is exact. It also becomes asymptotically exact in the limit $\Delta \to \pm\infty$. In the gapless XY regime, $|\Delta|<1$ under PBC, the $\mathrm{JW} \, \circ \, \Pi$ transformation consistently gives a lower sBCS energy than the other SFS transformations for other system sizes tested. At $\Delta=1$, all the SFS methods yield identical results, and $\mathrm{JW} \, \circ \, \Pi$ yields higher energy than the other beyond $\Delta=1$. This distinction is strongly suppressed under OBC, where the corresponding energy difference decreases with increasing system size. For example, for 12 sites, the energy curves resemble the PBC pattern.

Figure \ref{fig:xxz_2x4_pbc_obc_errors} shows the energy errors for the 2D SFS-transformed XXZ Hamiltonian. The results suggest we achieve better performance in 1D than in 2D due to the presence of strings in 2D. Similar to 1D, we see that under PBC in the XY-phase $\mathrm{JW} \, \circ \, \Pi$ yields energies superior to those of the other SFS-transformations, but yields energies higher than the others for $\Delta>1$. We also notice that beyond $\Delta \approx1.2$, the curves of both the labelings for $\mathrm{JW} \, \circ \, \Pi$ are very close to each other, with results for the $2\times4$ labeling being slightly better than those for the $4\times2$ labeling. Under OBC, we get frequent crossovers between the curves for $2\times4$ but a PBC-like pattern for $4\times2$.

\section{Beyond Mean-Field Theory}
In Sec.
\ref{SFS_MF_Results}, we showed that the mean-field of the SFS-transformed Hamiltonian yields promising results; clearly, however, there is room for improvement. For now, we will use coupled cluster (CC) theory \cite{Bartlett2007CoupledCluster,Shavitt2009ManyBody,Hagen2014NuclearCC,Bishop1991OverviewCC,Crawford2000IntroCC} to incorporate correlations atop sBCS. We will show results for the 1D XXZ model in OBC.

\begin{figure}[t]
\includegraphics[width=\columnwidth]{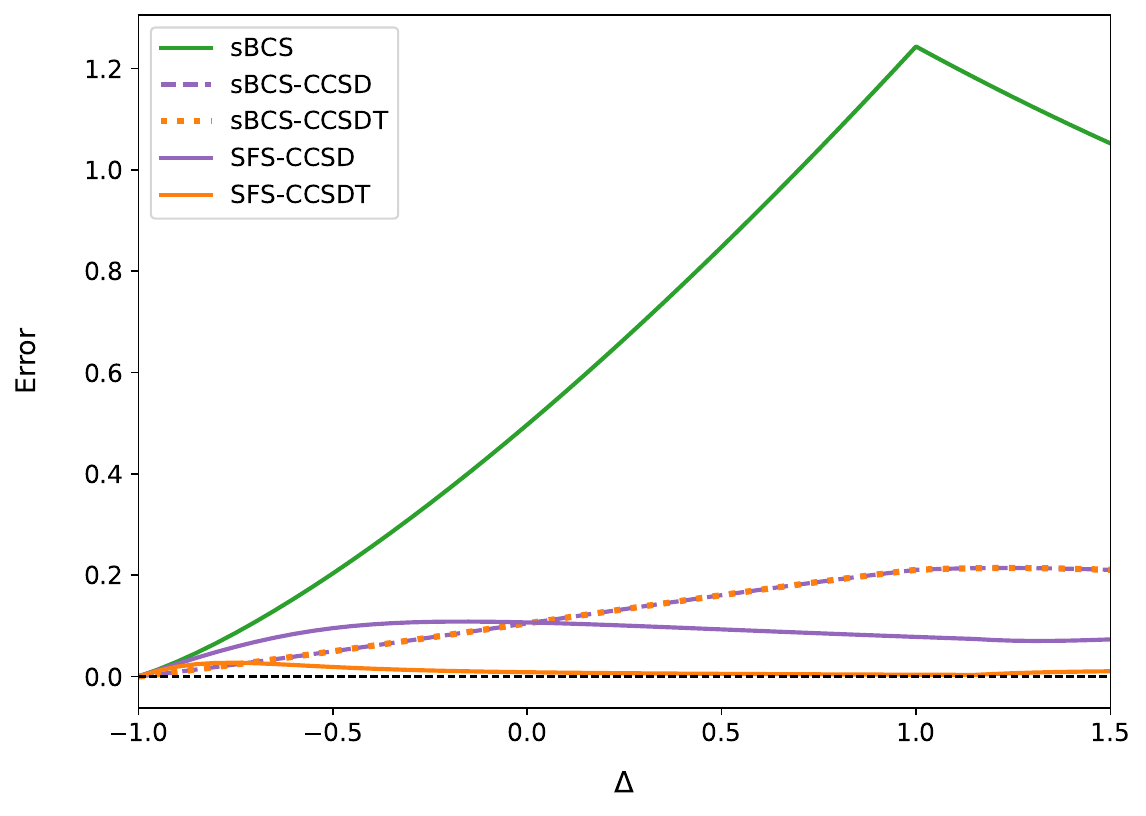}
\caption{Coupled-cluster energy errors for the 6-site XXZ Hamiltonian with OBC. SFS-CCSD and SFS-CCSDT use the optimized sBCS reference  obtained for $\Pi\circ\mathrm{JW}$-transformed Hamiltonian.}
\label{fig:sfs_cc}
\end{figure}

\subsection{Spin-BCS Coupled-Cluster Theory}
Let us begin with an overview of BCS coupled cluster theory.\cite{Henderson2014PRC} The correlated wavefunction is written as
\begin{equation}
\ket{\Psi} = \mathrm{e}^T \, \ket{\mathrm{sBCS}},
\end{equation}
where $T$ is the cluster operator that creates spin flips in the sBCS reference. We generically write $T$ as
\begin{equation}
T = \sum_Q t_Q \, \mathcal{S}_Q^+
\end{equation}
where  we define the excitation operator
\begin{equation}
\mathcal{S}_Q^{+}
=
\prod_{p\in Q}\mathcal{S}_p^{+},
\end{equation}  
which flips the spins on the sites contained
in \(Q\) relative to their local orientations in the sBCS
reference. The corresponding de-excitation operator is
$\mathcal{S}_Q^-=(\mathcal{S}_Q^+)^\dagger$.

When the sum over $Q$ runs over all possible configurations (i.e. all possible ways of flipping one spin, or of flipping two spins at a time, three spins at a time, and so on), this wavefunction with suitably chosen amplitudes $t_Q$ is exact so long as the sBCS state has overlap with the exact wavefunction.

To extract the energy and amplitudes, we insert the CC wavefunction into the Schr\"odinger equation and multiply on the left by $\mathrm{e}^{-T}$ to build a similarity-transformed Hamiltonian which we solve projectively:
\begin{subequations}
\begin{align}
E_{\mathrm{CC}} &= \bra{\mathrm{sBCS}} \mathrm{e}^{-T} \, H \, \mathrm{e}^T \ket{\mathrm{sBCS}},
\\
0 &= \bra{\mathrm{sBCS}} \mathcal{S}_Q^- \,  \mathrm{e}^{-T} \, H \, \mathrm{e}^T \ket{\mathrm{sBCS}}.
\end{align}
\end{subequations}
In practice, we restrict the summation over $Q$, usually by limiting the number of sites whose spins can be flipped at one time.  In coupled cluster doubles (CCD), we flip two spins at a time; coupled cluster singles and doubles (CCSD) flips one or two spins, coupled cluster with singles, doubles, and triples (CCSDT) flips one, two, or three, and so forth.

In terms of the computational scaling, standard sBCS based CCSD and CCSDT scales as $\mathcal{O}(M^3)$ and $\mathcal{O}(M^4)$ respectively. The scaling for the SFS-sBCS based CC remains the same, although the prefactor might vary depending on the contractions. 

\subsection{Results}
Generally speaking, the presence of strings like $\phi_p$ or $\psi_p$ renders the CC ansatz difficult to handle. We should emphasize that it is possible to handle strings by extending the disentangled cluster approximation used in projected coupled cluster theory,\cite{Qiu2017ProjectedCC,Qiu2019PNPBogoliubovCC,Qiu2018ProjectedCCOptimization,Song2022PowerSeriesSPCC} where we face a similar problem of having a string-like operator (in this case a symmetry rotation operator) acting on a coupled cluster wavefunction.

In the present calculations, we avoid the complications associated with nonlocal strings by limiting ourselves to the 1D XXZ case with open boundary conditions. The CC amplitude equations were solved using modified Broyden method,\cite{Qiu2018ProjectedCCOptimization,Johnson1988ModifiedBroyden,Baran2008BroydenNuclear} with convergence defined by a maximum residual smaller
than $10^{-10}$. The amplitude equations were obtained by using the
symbolic algebra package \textit{drudge} and its accompanying
automatic code generation facility \textit{gristmill}\cite{ZhaoScuseriaDrudgeGristmill}. Results for the 6-site model are shown in Fig. \ref{fig:sfs_cc}. For this, we report SFS-CCSD and SFS-CCSDT which refers to CCSD and CCSDT calculations based on sBCS obtained from $\Pi\circ\mathrm{JW}$ transformed 1D XXZ. The corresponding calculations based on the sBCS reference optimized
for the original (untransformed) XXZ Hamiltonian are denoted sBCS-CCSD and
sBCS-CCSDT. As expected, the inclusion of CC correlations on sBCS substantially improves the energy. In particular, we obtain more accurate results when applying CC to the SFS-transformed Hamiltonian than to the original Hamiltonian. Note that in the original representation, sBCS-CCSD and sBCS-CCSDT yield identical energies.  This is related to the extreme bimodal nature of the sBCS reference,\cite{Liu2023SpinAGP}, where $u_p = \pm v_p$, and the signs alternate from site to site. This additional structure causes odd excitations to vanish, essentially because the sBCS reference becomes an $S_x$ eigenstate for which odd excitations vanish by symmetry. By contrast, adding correlations on top of SFS sBCS with CCSD reduces the error substantially, and SFS-CCSDT closely tracks the exact energy over nearly the entire range. The systematic
improvement from SFS-sBCS to SFS-CCSD and SFS-CCSDT indicates that the SFS-transformed Hamiltonian can be accurately described at relatively low
excitation rank, consistent with the improved accuracy observed
at the mean-field level.

\section{Discussion}

Each SFS transformation considered here is an exact canonical transformation of the spin operator algebra and therefore preserves the exact spectrum of the original spin Hamiltonian. In other words, it represents the original Hamiltonian in a different basis. At the mean-field level, however, these transformations yield different results. This observation, along with the fact that solving spin Hamiltonians in other representations \cite{Henderson2022StrongWeak, Henderson2024FermionicMeanField} improves the mean-field, is the central motivation for the present work. The resulting mean-field state is expected to provide a more compact starting point for post-mean-field treatments, such as CC theory. The numerical results presented above suggest that solving the SFS-transformed Hamiltonian substantially improves the mean-field energy relative to the standard sBCS ansatz.

Each SFS transformation is a sequential mapping of JW and $\Pi$/$\Pi_R$ transformation; for example, we can first do the JW transformation and then the $\Pi$ transformation or vice versa. Other, more general, transformations can also be considered. However, we have observed in the past that the JW or parity transformation of spin systems to fermionic ones seems to yield the best accuracy, so one might suspect that SFS transformations involving these candidates are also likely to be superior. That said, more geometrically local transformations that minimize the strings might be of use for larger lattices.

In the 1D systems, where strings drop out when we order sites sequentially, our results are very encouraging, especially after correlation is included. Even in 1D, we could order the sites differently and might obtain different results; this is what we observe when we use the JW transformation to map spins to fermions and solve the Hamiltonian with fermionic mean-field\cite{Henderson2022StrongWeak,Henderson2024FermionicMeanField,RingSchuck1980,Valatin1961GHF,Schmid2004ProjectedHFB,JimenezHoyos2012ProjectedHF}, and we see no reason to expect anything else with the composite mappings.

Once the system becomes 2D, strings appear, and labeling effects clearly become significant. For the JW transformation, labeling effects can be removed by the use of more general JW strings,\cite{Henderson2023EJW} but we have not yet constructed similar generalizations of the strings for the parity mapping, and we need labeling invariance in both the mapping from spins to fermions and the mapping from fermions back to spins if we are to obtain final results independent of labeling. The above fact makes the generalization of strings for SFS transformation more involved, and thus we leave their construction for future
work.

While the SFS transformations we have considered have provided new Hamiltonians whose mean-field solution is more accurate than is the mean-field solution of the original Hamiltonian, this is by no means guaranteed to occur.  After all, one could start with the XXZ Hamiltonian after the SFS mapping (cf. Eq. \ref{Eqn:XXZHamiltonianMapped}), do the SFS mapping in the other direction, and obtain the more familiar XXZ model of Eq. \ref{Eqn:XXZHamiltonianOriginal}.  But when the SFS transformation provides superior mean-field results, we would generally expect it to provide superior correlated results as well, supposing for the moment that one can actually do the correlated calculation with a more complicated Hamiltonian that may be higher-body or may have strings.

Finally, we note that in general, different SFS mappings may yield different results and some could be superior to others for a given magnetic phase.  To the extent that different SFS orderings are favored in different magnetic regimes, the choice of nonlinear canonical transformation may be used as an additional degree of freedom in constructing compact many-body approximations.  Regardless, we suggest that nonlinear canonical transformations may provide a promising route toward more compact descriptions of strongly-correlated spin systems.

\section*{Supplementary Material}

See the supplementary material for a finite-size analysis of the
$J_1$--$J_2$ model on square lattices with periodic boundary
conditions. We compare the energies per site obtained with
conventional sBCS and SFS-sBCS, and examine the size-extensivity of the mean-field obtained from the SFS transformation.

\section*{Data Availability}
The data that support the findings of this study are available from the corresponding author upon reasonable request.

\section*{Author Declaration}

\subsection*{Conflict of interest}
The authors have no conflicts to disclose.
\subsection*{Author contribution}
\textbf{S. Ghosh}: Conceptualization (equal); Data curation (lead);
Formal analysis (lead); Investigation (lead); Methodology
(equal); Software (lead); Validation (lead); Visualization
(lead); Writing -- original draft (lead); Writing -- review
\& editing (equal).

\textbf{T. M. Henderson}: Conceptualization (equal); Formal analysis
(equal); Methodology (equal); Supervision (equal); Validation
(equal); Writing -- review \& editing (equal).

\textbf{G. E. Scuseria}: Conceptualization (equal); Funding acquisition
(lead); Methodology (equal); Project administration (lead);
Resources (lead); Supervision (lead); Writing -- review \&
editing (equal).
\begin{acknowledgments}
This work was supported by the U.S. Department of Energy, Office of Basic Energy Sciences, under Award DE-SC0001474.  G.E.S. is a Welch Foundation Chair (C-0036).
\end{acknowledgments}

\appendix

\section{Expressions for various other canonical transformations}
\label{sfs_expressions}

In this Appendix, we collect the explicit operator transformations used in
this work. We use the convention \(A\circ B\) which means that
\(B\) is applied first, followed by \(A\).

The transformation $\Pi_R \circ \text{JW}$ maps the Paulis as 

\begin{subequations}
\begin{align}
\sigma^x_p &\mathrel{\underset{\scriptsize{\Pi}_R \circ \text{JW}}{\mapsto}}  \psi_p^z =\prod_{q\ge p} \sigma^z_q,\\
\sigma^y_p &\mathrel{\underset{\scriptsize{\Pi}_R\circ \text{JW}}{\mapsto}} -\text{i}\,\psi_p^z\,\sigma^y_{p-1}\sigma^y_p,\\
\sigma^z_p &\mathrel{\underset{\scriptsize{\Pi}_R\circ \text{JW}}{\mapsto}} -\sigma^y_{p-1}\sigma^y_p,
\end{align}
\end{subequations}

The transformation \(\mathrm{JW}\circ\Pi\) maps the local Pauli operators as

\begin{subequations}
    \begin{align}
        \sigma^x_p &\mathrel{\underset{\scriptsize{\text{JW}\circ \Pi}}{\mapsto}} \sigma^x_{p}\sigma^x_{p+1}\\
        \sigma^y_p &\mathrel{\underset{\scriptsize{\text{JW}\circ \Pi}}{\mapsto}} -\,\tilde{\phi_p}\,\sigma^x_{p+1}\sigma^y_{p}\\
        \sigma^z_p &\mathrel{\underset{\scriptsize{\text{JW}\circ \Pi}}{\mapsto}} -\tilde{\phi_p}\, \sigma^z_p\\
        \tilde{\phi}_p & = \prod_{q<p} (-\sigma^z_q)
    \end{align}
\end{subequations}

The transformation \(\mathrm{JW}\circ\Pi_R\) maps the local Pauli
operators as
\begin{subequations}
    \begin{align}
        \sigma^x_p &\mathrel{\underset{\scriptsize{\text{JW}\circ \Pi_R}}{\mapsto}} -\tilde{\phi}_p\,\sigma^x_{p+1}\sigma^y_p\\
        \sigma^y_p &\mathrel{\underset{\scriptsize{\text{JW}\circ \Pi_R}}{\mapsto}} -\,\tilde{\phi_p}\,\sigma^z_p\\
        \sigma^z_p &\mathrel{\underset{\scriptsize{\text{JW}\circ \Pi_R}}{\mapsto}} \sigma^x_p\sigma^x_{p+1}\\
        \tilde{\phi}_p  &= \prod_{q<p} (-\sigma^z_q)
    \end{align}
\end{subequations}

Additional care is required when applying these transformations near the boundaries. For example,
with 1D OBC, \(p=1,\ldots,M-1\) in the
bond Hamiltonian, and we take
\[
\sigma_0^\alpha \equiv 1,
\qquad
\sigma_{M+1}^x \equiv 1,
\]
where $\alpha\in \{x,y,z\}$.
For PBC, the \(M\) to \(1\) boundary bond is
treated separately, since the SFS transformation produces additional
nonlocal strings across the boundary.

\section{Circuit representation of the
$\Pi\circ\mathrm{JW}$ transformation}
\label{App:SFSCircuit}

In this Appendix, we derive the unitary circuit that implements
the composite $\Pi\circ\mathrm{JW}$ transformation. We use the
abbreviations
\begin{equation}
X_p\equiv\sigma_p^x,
\qquad
Y_p\equiv\sigma_p^y,
\qquad
Z_p\equiv\sigma_p^z,
\end{equation}
and adopt the boundary convention
\begin{equation}
Z_0\equiv\mathbb{I}.
\end{equation}

\subsection{CNOT staircase}

Let
\begin{equation}
C_p
=
\operatorname{CNOT}_{p\rightarrow p+1}
\end{equation}
denote a controlled-NOT gate with site $p$ as the control and
site $p+1$ as the target. In the standard computational basis,
\begin{equation}
C_p
|a\rangle_p|b\rangle_{p+1}
=
|a\rangle_p|a\oplus b\rangle_{p+1},
\end{equation}
where $a,b\in\{0,1\}$ and $\oplus$ denotes addition modulo two. Here, the computational-basis states are identified with the
local spin states according to
$|0\rangle\equiv|\uparrow\rangle$ and
$|1\rangle\equiv|\downarrow\rangle$. These computational-basis labels are distinct from the fermionic occupation labels used in Sec. \ref{Spin-Fermion Transformations}.
The CNOT gate is Hermitian and involutory,
\begin{equation}
C_p^\dagger=C_p,
\qquad
C_p^2=\mathbb{I}.
\end{equation}
Its action on the Pauli generators under conjugation is
\begin{subequations}
\label{Eqn:CNOTPauliRules}
\begin{align}
C_pX_pC_p^\dagger
&=
X_pX_{p+1},
\label{Eqn:CNOTXControl}
\\
C_pX_{p+1}C_p^\dagger
&=
X_{p+1},
\label{Eqn:CNOTXTarget}
\\
C_pZ_pC_p^\dagger
&=
Z_p,
\label{Eqn:CNOTZControl}
\\
C_pZ_{p+1}C_p^\dagger
&=
Z_pZ_{p+1}.
\label{Eqn:CNOTZTarget}
\end{align}
\end{subequations}
Operators on all other sites are unchanged. Because conjugation
by $C_p$ maps Pauli operators to Pauli products, $C_p$ is a
Clifford gate.

For a system of $M$ sites, define the ordered staircase
\begin{align}
U_{\mathrm C}
&=
C_{M-1}C_{M-2}\cdots C_2C_1
\nonumber\\
&=
\operatorname{CNOT}_{M-1\rightarrow M}
\cdots
\operatorname{CNOT}_{2\rightarrow3}
\operatorname{CNOT}_{1\rightarrow2},
\label{Eqn:AppendixCNOTStaircase}
\end{align}
where the rightmost gate acts first. The circuit therefore
contains $M-1$ CNOT gates connecting consecutive sites in the
chosen one-dimensional ordering. For a sequentially labeled
one-dimensional chain, these sites are also geometrical nearest
neighbors.

\subsection{Action of the staircase on the Pauli generators}

Consider first a local $X_p$ operator. The gates $C_j$ with
$j<p$ leave $X_p$ unchanged. In particular, $C_{p-1}$ treats
site $p$ as its target, and an $X$ operator on the target is
unchanged according to Eq.~\eqref{Eqn:CNOTXTarget}. The first
nontrivial action is therefore due to $C_p$:
\begin{equation}
X_p
\xrightarrow{C_p}
X_pX_{p+1}.
\end{equation}
Each subsequent gate extends the Pauli string by one site,
\begin{equation}
X_p
\xrightarrow{C_p}
X_pX_{p+1}
\xrightarrow{C_{p+1}}
X_pX_{p+1}X_{p+2}
\xrightarrow{}
\cdots
\xrightarrow{C_{M-1}}
\prod_{q=p}^{M}X_q.
\end{equation}
Consequently,
\begin{equation}
U_{\mathrm C}X_pU_{\mathrm C}^\dagger
=
\prod_{q=p}^{M}X_q.
\label{Eqn:AppendixStaircaseX}
\end{equation}

For $p>1$, the first gate that acts nontrivially on $Z_p$ is
$C_{p-1}$, for which site $p$ is the target. Equation
\eqref{Eqn:CNOTZTarget} gives
\begin{equation}
Z_p
\xrightarrow{C_{p-1}}
Z_{p-1}Z_p.
\end{equation}
The next gate, $C_p$, treats site $p$ as its control and
therefore leaves $Z_p$ unchanged. All later gates
$C_{p+1},\ldots,C_{M-1}$ act strictly to the right of site $p$
and leave the product unchanged. Hence,
\begin{equation}
U_{\mathrm C}Z_pU_{\mathrm C}^\dagger
=
Z_{p-1}Z_p,
\qquad p>1.
\end{equation}
For $p=1$, no CNOT has site $1$ as its target, and therefore
\begin{equation}
U_{\mathrm C}Z_1U_{\mathrm C}^\dagger
=
Z_1.
\end{equation}
Using $Z_0=\mathbb{I}$, both cases may be written as
\begin{equation}
U_{\mathrm C}Z_pU_{\mathrm C}^\dagger
=
Z_{p-1}Z_p.
\label{Eqn:AppendixStaircaseZ}
\end{equation}

Thus, the CNOT staircase generates the string and bond operators
appearing in the SFS transformation, apart from the overall sign
in the transformation of $Z_p$.

\subsection{Complete SFS unitary}

To reproduce the sign convention of the
$\Pi\circ\mathrm{JW}$ transformation, define a layer of local
spin flips,
\begin{equation}
U_X
=
\prod_{r=1}^{M}X_r.
\label{Eqn:AppendixUX}
\end{equation}
Because Pauli operators on different sites commute,
\begin{subequations}
\begin{align}
U_XX_pU_X^\dagger
&=
X_p,
\\
U_XZ_pU_X^\dagger
&=
X_pZ_pX_p
=
-Z_p,
\\
U_XY_pU_X^\dagger
&=
X_pY_pX_p
=
-Y_p.
\end{align}
\end{subequations}
The complete unitary is therefore
\begin{equation}
U_{\Pi\circ\mathrm{JW}}
=
U_{\mathrm C}U_X.
\label{Eqn:AppendixSFSUnitary}
\end{equation}
Since operators act from right to left, the local $X$ layer is
applied first, followed by the CNOT staircase.

Using Eqs.~\eqref{Eqn:AppendixStaircaseX} and
\eqref{Eqn:AppendixStaircaseZ}, the transformed $X_p$ operator is
\begin{align}
U_{\Pi\circ\mathrm{JW}}
X_p
U_{\Pi\circ\mathrm{JW}}^\dagger
&=
U_{\mathrm C}
\left(
U_XX_pU_X^\dagger
\right)
U_{\mathrm C}^\dagger
\nonumber\\
&=
\prod_{q=p}^{M}X_q.
\label{Eqn:AppendixSFSX}
\end{align}
Similarly,
\begin{align}
U_{\Pi\circ\mathrm{JW}}
Z_p
U_{\Pi\circ\mathrm{JW}}^\dagger
&=
U_{\mathrm C}
\left(
U_XZ_pU_X^\dagger
\right)
U_{\mathrm C}^\dagger
\nonumber\\
&=
-Z_{p-1}Z_p.
\label{Eqn:AppendixSFSZ}
\end{align}

The transformation of $Y_p$ follows from
\begin{equation}
Y_p=\mathrm{i}X_pZ_p.
\end{equation}
Since unitary conjugation preserves operator products,
\begin{align}
U_{\Pi\circ\mathrm{JW}}
Y_p
U_{\Pi\circ\mathrm{JW}}^\dagger
&=
\mathrm{i}
\left(
U_{\Pi\circ\mathrm{JW}}
X_p
U_{\Pi\circ\mathrm{JW}}^\dagger
\right)
\left(
U_{\Pi\circ\mathrm{JW}}
Z_p
U_{\Pi\circ\mathrm{JW}}^\dagger
\right)
\nonumber\\
&=
-\mathrm{i}
\left(
\prod_{q=p}^{M}X_q
\right)
Z_{p-1}Z_p.
\label{Eqn:AppendixSFSY}
\end{align}

Restoring the original Pauli notation, the complete circuit
action is
\begin{subequations}
\label{Eqn:AppendixCompleteSFSMapping}
\begin{align}
U_{\Pi\circ\mathrm{JW}}\,
\sigma_p^x\,
U_{\Pi\circ\mathrm{JW}}^\dagger
&=
\prod_{q=p}^{M}\sigma_q^x,
\\
U_{\Pi\circ\mathrm{JW}}\,
\sigma_p^y\,
U_{\Pi\circ\mathrm{JW}}^\dagger
&=
-\mathrm{i}
\left(
\prod_{q=p}^{M}\sigma_q^x
\right)
\sigma_{p-1}^z\sigma_p^z,
\\
U_{\Pi\circ\mathrm{JW}}\,
\sigma_p^z\,
U_{\Pi\circ\mathrm{JW}}^\dagger
&=
-\sigma_{p-1}^z\sigma_p^z.
\end{align}
\end{subequations}
These relations coincide with those obtained by composing the
Jordan--Wigner and parity transformations directly.

The initial spin-flip layer may equivalently be moved through
the CNOT staircase. From Eq.~\eqref{Eqn:AppendixStaircaseX},
\begin{align}
U_{\mathrm C}U_XU_{\mathrm C}^\dagger
&=
\prod_{p=1}^{M}
\left(
U_{\mathrm C}X_pU_{\mathrm C}^\dagger
\right)
\nonumber\\
&=
\prod_{p=1}^{M}
\prod_{q=p}^{M}X_q.
\end{align}
For fixed $q$, the operator $X_q$ occurs once for each $p\le q$
and therefore occurs $q$ times. Since $X_q^2=\mathbb{1}$,
\begin{equation}
U_{\mathrm C}U_XU_{\mathrm C}^\dagger
=
\prod_{\substack{q=1\\q\text{ odd}}}^{M}X_q.
\end{equation}
It follows that
\begin{equation}
U_{\Pi\circ\mathrm{JW}}
=
\left(
\prod_{\substack{q=1\\q\text{ odd}}}^{M}X_q
\right)
U_{\mathrm C}.
\label{Eqn:AppendixAlternatingX}
\end{equation}
Thus, the same transformation may be implemented either by
applying $X$ to every site before the staircase or by applying
$X$ to alternating sites after the staircase.

Finally, the unitary representation guarantees preservation of
the operator algebra. For arbitrary operators $A$ and $B$,
\begin{subequations}
\begin{align}
\left(UAU^\dagger\right)
\left(UBU^\dagger\right)
&=
UABU^\dagger,
\\
\left[
UAU^\dagger,
UBU^\dagger
\right]
&=
U[A,B]U^\dagger.
\end{align}
\end{subequations}
The transformed generators therefore commute on different sites
and satisfy the same on-site $\mathfrak{su}(2)$ algebra as the
original Pauli operators. For $p>1$,
$-\sigma_{p-1}^z\sigma_p^z$ is a bond or a domain-wall variable,
whereas $\sigma_p^x$ is mapped to an extended string. For
$p=1$, the convention $\sigma_0^z=1$ instead produces
a boundary variable.

\section{Computational scaling of SFS mean-field theory}
\label{app:sfs_scaling}

We show that the SFS transformation does not increase the
asymptotic cost of mean-field energy evaluation when
intermediate string products are reused.

Consider a spin Hamiltonian containing $K$ nonzero
two-body interaction terms,
\begin{equation}
H =
\sum_{(p,q)}
\sum_{\alpha\beta}
J_{pq}^{\alpha\beta}
\sigma_p^\alpha \sigma_q^\beta.
\end{equation}
The sBCS wavefunction is a product of local spinors,
\begin{equation}
|\Phi\rangle =
\bigotimes_{p=1}^{M}
\left(
u_p|\downarrow\rangle_p+
v_p|\uparrow\rangle_p
\right).
\end{equation}
Consequently, expectation values of operators acting
on distinct sites factorize,
\begin{equation}
\langle\Phi|
\sigma_p^\alpha\sigma_q^\beta
|\Phi\rangle
=
\langle\sigma_p^\alpha\rangle
\langle\sigma_q^\beta\rangle.
\end{equation}
After $\mathcal{O}(M)$ preprocessing of the local expectation
values, each interaction can be evaluated in constant
time. Conventional mean-field energy evaluation
therefore requires $\mathcal{O}(M+K)$ operations.

The SFS transformation maps local operators into Pauli
products, which may include nonlocal strings. For
example, under the $\Pi\circ\mathrm{JW}$ transformation,
\begin{equation}
\sigma_p^x \longmapsto
\psi_p =
\prod_{k=p}^{M}\sigma_k^x.
\end{equation}
For $p<q$, the product of two such operators simplifies to
\begin{equation}
\psi_p\psi_q =
\prod_{k=p}^{q-1}\sigma_k^x,
\end{equation}
where we have used $(\sigma_k^x)^2=1$.
Its expectation value factorizes as
\begin{equation}
\langle\Phi|\psi_p\psi_q|\Phi\rangle
=
\prod_{k=p}^{q-1}x_k,
\qquad
x_k =
\langle\Phi|\sigma_k^x|\Phi\rangle.
\end{equation}

To evaluate these products efficiently, define the
prefix products
\begin{equation}
F_j=\prod_{k=1}^{j}x_k,
\qquad F_0=1.
\end{equation}
All $F_j$ can be constructed in $\mathcal{O}(M)$ operations.
When the denominator is nonzero, an arbitrary interval
product is then obtained as
\begin{equation}
\prod_{k=p}^{q-1}x_k
=
\frac{F_{q-1}}{F_{p-1}},
\end{equation}
which requires $\mathcal{O}(1)$ operations. Vanishing factors can
be accommodated by separately tracking the number of
zeros and the product of nonzero factors within each
interval.

The same factorization can be understood in the
quasispin basis introduced in Sec.~IV. For a local
Pauli operator $O_k$, we may write
\begin{equation}
O_k|\downarrow_k'\rangle
=
A_k|\downarrow_k'\rangle
+
B_k|\uparrow_k'\rangle,
\end{equation}
where $A_k=\langle\downarrow_k'|O_k|
\downarrow_k'\rangle$. The interior of a string
therefore contributes a product of the $A_k$ factors. At a site containing an additional endpoint operator,
the corresponding factor is replaced by a local
contraction, for example,
\begin{equation}
\langle\downarrow_k'|
(a_k+b_k\mathcal{S}_k^-)
(A_k+B_k\mathcal{S}_k^+)
|\downarrow_k'\rangle
=
a_kA_k+b_kB_k.
\end{equation}
Since the number of endpoint sites is bounded,
these additional contractions require only a
constant number of operations.

More generally, the transformed two-body interactions
considered here contain a bounded number of local
endpoint operators together with Pauli strings supported
on a bounded number of intervals. Since the sBCS
wavefunction factorizes, the corresponding matrix
elements reduce to products of local expectation values.
The interval contributions can be obtained from the
precomputed products, while the remaining endpoint
contractions require only a constant number of
operations. Hence, each transformed interaction can be evaluated in $\mathcal{O}(1)$ time after $\mathcal{O}(M)$ preprocessing.

Because the SFS transformation maps each Pauli product
to another Pauli product, the number of terms generated per original interaction remains bounded. The total
cost of an efficiently implemented SFS mean-field
energy evaluation is therefore
\begin{equation}
T_{\mathrm{SFS-MF}}=\mathcal{O}(M+K).
\end{equation}
For finite-range spin Hamiltonians,
this gives $\mathcal{O}(M)$ scaling, whereas a generic two-body
Hamiltonian with $K=\mathcal{O}(M^2)$ requires $\mathcal{O}(M^2)$
operations. These are the same asymptotic exponents as
conventional sBCS mean-field theory. The additional
strings, in general,  increase the computational prefactor without changing the leading scaling exponent.

\bibliographystyle{aipnum4-2}
\bibliography{sfs_refs}

@article{BravyiKitaev2002,
  author  = {Bravyi, Sergey B. and Kitaev, Alexei Yu.},
  title   = {Fermionic quantum computation},
  journal = {Annals of Physics},
  volume  = 298,
  pages   = {210},
  year    = {2002}
   
}

@article{Liu2023SpinAGP,
  title = {Exploring Spin Antisymmetrized Geminal Power Ans{\"a}tze for Strongly Correlated Spin Systems},
  author = {Liu, Z. and Gao, F. and Chen, G. P. and Henderson, T. M. and Dukelsky, J. and Scuseria, G. E.},
  journal = {Physical Review B},
  volume = {108},
  number = {8},
  pages = {085136},
  year = {2023},
  doi = {10.1103/PhysRevB.108.085136},
}

@article{JordanWigner1928,
  author  = {Jordan, Pascual and Wigner, Eugene},
  title   = {{\"U}ber das Paulische {\"A}quivalenzverbot},
  journal = {Zeitschrift f{\"u}r Physik},
  volume  = {47},
  pages   = {631--651},
  year    = {1928}
}

@article{Chien2026SimulatingFermions,
  author  = {Chien, Riley W. and Chiew, Mitchell L. and Harrison, Brent and Necaise, Jason and Wang, Weishi and Mudassar, Maryam and McLauchlan, Campbell and Henderson, Thomas M. and Scuseria, Gustavo E. and Strelchuk, Sergii and Whitfield, James D.},
  title   = {Simulating fermions with a digital quantum computer},
  journal = {Nature Reviews Physics},
  volume  = {8},
  pages   = {131--145},
  year    = {2026},
  doi     = {10.1038/s42254-025-00914-5}
}

@article{Henderson2022StrongWeak,
  author  = {Henderson, Thomas M. and Chen, Guo P. and Scuseria, Gustavo E.},
  title   = {Strong--weak duality via {Jordan--Wigner} transformation: Using fermionic methods for strongly correlated {su(2)} spin systems},
  journal = {The Journal of Chemical Physics},
  volume  = {157},
  pages   = {194114},
  year    = {2022},
  doi     = {10.1063/5.0125124}
}

@article{Wang2013GaplessSpinLiquidJ1J2,
  author  = {Wang, Ling and Poilblanc, Didier and Gu, Zheng-Cheng and Wen, Xiao-Gang and Verstraete, Frank},
  title   = {Constructing a gapless spin-liquid state for the spin-$1/2$ {$J_1$-$J_2$} {Heisenberg} model on a square lattice},
  journal = {Physical Review Letters},
  volume  = {111},
  pages   = {037202},
  year    = {2013},
  doi     = {10.1103/PhysRevLett.111.037202}
}

@article{Henderson2023EJW,
  author  = {Henderson, Thomas M. and Gao, Fei and Scuseria, Gustavo E.},
  title   = {Restoring permutational invariance in the Jordan--Wigner transformation},
  journal = {Molecular Physics},
  volume  = {122},
  number  = {7--8},
  year    = {2023},
  doi     = {10.1080/00268976.2023.2254857}
}

@article{Darradi2008,
  author = {Darradi, R. and Derzhko, O. and Zinke, R. and Schulenburg, J. and Kr{\"u}ger, S. E. and Richter, J.},
  title = {},
  journal = {Phys. Rev. B},
  volume = {78},
  pages = {214415},
  year = {2008},
  doi  = {https://doi.org/10.1103/PhysRevB.78.214415}
}

@article{Gong2014,
  author = {Gong, S.-S. and Zhu, W. and Sheng, D. N. and Motrunich, O. I. and Fisher, M. P. A.},
  title = {},
  journal = {Phys. Rev. Lett.},
  volume = {113},
  pages = {027201},
  year = {2014},
  doi  = {https://doi.org/10.1103/PhysRevLett.113.027201}
}

@article{Richter2015,
  author = {Richter, J. and Zinke, R. and Farnell, D. J. J.},
  title = {},
  journal = {Eur. Phys. J. B},
  volume = {88},
  pages = {2},
  year = {2015},
  doi  = {https://doi.org/10.1140/epjb/e2014-50589-x

}
}

@article{Henderson2024FermionicMeanField,
  author  = {Henderson, Thomas M. and Harrison, Brent and Magoulas, Ilias and Necaise, Jason and Projansky, Andrew M. and Evangelista, Francesco A. and Whitfield, James D. and Scuseria, Gustavo E.},
  title   = {Fermionic mean-field theory as a tool for studying spin Hamiltonians},
  journal = {The Journal of Chemical Physics},
  volume  = {161},
  pages   = {234112},
  year    = {2024},
  doi     = {10.1063/5.0242219}
}

@article{Majumdar1969,
  author = {Majumdar, C. and Ghosh, D. K.},
  title = {},
  journal = {J. Math. Phys.},
  volume = {10},
  pages = {1388},
  year = {1969}
}

@article{Amico2008Entanglement,
  author  = {Amico, Luigi and Fazio, Rosario and Osterloh, Andreas and Vedral, Vlatko},
  title   = {Entanglement in many-body systems},
  journal = {Reviews of Modern Physics},
  volume  = {80},
  pages   = {517--576},
  year    = {2008},
  doi     = {10.1103/RevModPhys.80.517}
}

@article{Eisert2010AreaLaws,
  author  = {Eisert, Jens and Cramer, Marcus and Plenio, Martin B.},
  title   = {Colloquium: Area laws for the entanglement entropy},
  journal = {Reviews of Modern Physics},
  volume  = {82},
  pages   = {277--306},
  year    = {2010},
  doi     = {10.1103/RevModPhys.82.277}
}

@article{Schollwoeck2011DMRG,
  author  = {Schollw{\"o}ck, Ulrich},
  title   = {The density-matrix renormalization group in the age of matrix product states},
  journal = {Annals of Physics},
  volume  = {326},
  pages   = {96--192},
  year    = {2011},
  doi     = {10.1016/j.aop.2010.09.012}
}

@book{RingSchuck1980,
  author    = {Ring, P. and Schuck, P.},
  title     = {The Nuclear Many-Body Problem},
  publisher = {Springer-Verlag},
  address   = {New York},
  year      = {1980}
}

@article{Seeley2012BK,
  author  = {Seeley, Jacob T. and Richard, Martin J. and Love, Peter J.},
  title   = {The Bravyi-Kitaev transformation for quantum computation of electronic structure},
  journal = {The Journal of Chemical Physics},
  volume  = {137},
  number  = {22},
  pages   = {224109},
  year    = {2012},
  doi     = {10.1063/1.4768229}
}

@incollection{Farnell2004CCMQuantumMagnetism,
  author    = {Farnell, D. J. J. and Bishop, R. F.},
  title     = {The coupled cluster method applied to quantum magnetism},
  booktitle = {Quantum Magnetism},
  editor    = {Schollw{\"o}ck, U. and Richter, J. and Farnell, D. J. J. and Bishop, R. F.},
  publisher = {Springer},
  address   = {Berlin, Heidelberg},
  year      = {2004},
  pages     = {307--348}
}

@article{Qiu2017ProjectedCC,
  author  = {Qiu, Yiheng and Henderson, Thomas M. and Zhao, Jinmo and Scuseria, Gustavo E.},
  title   = {Projected coupled cluster theory},
  journal = {The Journal of Chemical Physics},
  volume  = {147},
  pages   = {064111},
  year    = {2017},
  doi     = {10.1063/1.4991020}
}

@article{Henderson2014PRC,
  author  = {Henderson, Thomas M. and Dukelsky, Jorge and Signoracci, Angelo and Duguet, Thomas and Scuseria, Gustavo E.},
  title   = {Quasiparticle coupled cluster theory for pairing interactions},
  journal = {Physical Review C},
  volume  = {89},
  number  = {5},
  pages   = {054305},
  year    = {2014},
  doi     = {10.1103/PhysRevC.89.054305}
}

@article{Chen2023RobustWick,
  author  = {Chen, Guo P. and Scuseria, Gustavo E.},
  title   = {Robust formulation of {Wick's} theorem for computing matrix elements between {Hartree--Fock--Bogoliubov} wavefunctions},
  journal = {The Journal of Chemical Physics},
  volume  = {158},
  pages   = {231102},
  year    = {2023},
  doi     = {10.1063/5.0156124}
}

@article{Balian1969NonunitaryBogoliubov,
  author  = {Balian, R. and Br{\'e}zin, E.},
  title   = {Nonunitary {Bogoliubov} transformations and extension of {Wick's} theorem},
  journal = {Il Nuovo Cimento B},
  volume  = {64},
  pages   = {37--55},
  year    = {1969},
  doi     = {10.1007/BF02710281}
}

@article{Bishop1998PhaseTransitionsJ1J2,
author  = {Bishop, R. F. and Farnell, D. J. J. and Parkinson, J. B.},
title   = {Phase transitions in the spin-half {$J_1$--$J_2$} model},
journal = {Physical Review B},
volume  = {58},
pages   = {6394--6402},
year    = {1998},
doi     = {10.1103/PhysRevB.58.6394}
}

@article{Qiu2019PNPBogoliubovCC,
author  = {Qiu, Y. and Henderson, T. M. and Duguet, T. and Scuseria, G. E.},
title   = {Particle-number projected {Bogoliubov-coupled-cluster} theory: Application to the pairing Hamiltonian},
journal = {Physical Review C},
volume  = {99},
pages   = {044301},
year    = {2019},
doi     = {10.1103/PhysRevC.99.044301}
}

@article{Qiu2018ProjectedCCOptimization,
author  = {Qiu, Yiheng and Henderson, Thomas M. and Zhao, Jinmo and Scuseria, Gustavo E.},
title   = {Projected coupled cluster theory: Optimization of cluster amplitudes in the presence of symmetry projection},
journal = {The Journal of Chemical Physics},
volume  = {149},
pages   = {164108},
year    = {2018},
doi     = {10.1063/1.5053605},
eprint  = {1808.07972},
archivePrefix = {arXiv},
primaryClass  = {physics.chem-ph}
}

@article{Bartlett2007CoupledCluster,
author  = {Bartlett, Rodney J. and Musia{\l}, Monika},
title   = {Coupled-cluster theory in quantum chemistry},
journal = {Reviews of Modern Physics},
volume  = {79},
pages   = {291--352},
year    = {2007},
doi     = {10.1103/RevModPhys.79.291}
}

@book{Shavitt2009ManyBody,
author    = {Shavitt, Isaiah and Bartlett, Rodney J.},
title     = {Many-Body Methods in Chemistry and Physics: {MBPT} and Coupled-Cluster Theory},
publisher = {Cambridge University Press},
address   = {New York},
year      = {2009}
}

@article{Song2022PowerSeriesSPCC,
  author  = {Song, Ruiheng and Henderson, Thomas M. and Scuseria, Gustavo E.},
  title   = {A power series approximation in symmetry projected coupled cluster theory},
  journal = {The Journal of Chemical Physics},
  volume  = {156},
  number  = {10},
  pages   = {104105},
  year    = {2022},
  doi     = {10.1063/5.0080165}
}

@article{Hagen2014NuclearCC,
  author  = {Hagen, G. and Papenbrock, T. and Hjorth-Jensen, M. and Dean, D. J.},
  title   = {Coupled-cluster computations of atomic nuclei},
  journal = {Reports on Progress in Physics},
  volume  = {77},
  number  = {9},
  pages   = {096302},
  year    = {2014},
  doi     = {10.1088/0034-4885/77/9/096302}
}

@article{Bishop1991OverviewCC,
  author  = {Bishop, R. F.},
  title   = {An overview of coupled cluster theory and its applications in physics},
  journal = {Theoretica Chimica Acta},
  volume  = {80},
  pages   = {95--148},
  year    = {1991},
  doi     = {10.1007/BF01119617}
}

@incollection{Crawford2000IntroCC,
  author    = {Crawford, T. Daniel and Schaefer, Henry F.},
  title     = {An Introduction to Coupled Cluster Theory for Computational Chemists},
  booktitle = {Reviews in Computational Chemistry},
  volume    = {14},
  pages     = {33--136},
  year      = {2000},
  doi       = {10.1002/9780470125915.ch2}
}

@article{Yang1966XXZ2,
  author  = {Yang, C. N. and Yang, C. P.},
  title   = {One-Dimensional Chain of Anisotropic Spin-Spin Interactions. II. Properties of the Ground-State Energy per Lattice Site for an Infinite System},
  journal = {Physical Review},
  volume  = {150},
  pages   = {327--339},
  year    = {1966},
  doi     = {10.1103/PhysRev.150.327}
}

@article{Yang1966XXZ3,
  author  = {Yang, C. N. and Yang, C. P.},
  title   = {One-Dimensional Chain of Anisotropic Spin-Spin Interactions. III. Applications},
  journal = {Physical Review},
  volume  = {151},
  pages   = {258--264},
  year    = {1966},
  doi     = {10.1103/PhysRev.151.258}
}

@book{Giamarchi2004OneDimension,
  author    = {Giamarchi, Thierry},
  title     = {Quantum Physics in One Dimension},
  publisher = {Oxford University Press},
  address   = {Oxford},
  year      = {2004}
}

@article{Jiang2012SpinLiquidJ1J2,
  author  = {Jiang, Hong-Chen and Yao, Hong and Balents, Leon},
  title   = {Spin liquid ground state of the spin-$1/2$ square {$J_1$-$J_2$} {Heisenberg} model},
  journal = {Physical Review B},
  volume  = {86},
  pages   = {024424},
  year    = {2012},
  doi     = {10.1103/PhysRevB.86.024424}
}

@article{Murg2009PEPSFrustrated,
  author  = {Murg, V. and Verstraete, F. and Cirac, J. I.},
  title   = {Exploring frustrated spin systems using projected entangled pair states},
  journal = {Physical Review B},
  volume  = {79},
  pages   = {195119},
  year    = {2009},
  doi     = {10.1103/PhysRevB.79.195119}
}

@article{Yu2012PlaquetteTensorJ1J2,
  author  = {Yu, Ji-Feng and Kao, Ying-Jer},
  title   = {Spin-$1/2$ {$J_1$-$J_2$} {Heisenberg} antiferromagnet on a square lattice: A plaquette renormalized tensor network study},
  journal = {Physical Review B},
  volume  = {85},
  pages   = {094407},
  year    = {2012},
  doi     = {10.1103/PhysRevB.85.094407}
}

@article{Bethe1931BetheAnsatz,
author = {Bethe, H.},
title = {Zur Theorie der Metalle. I. Eigenwerte und Eigenfunktionen der linearen Atomkette},
journal = {Zeitschrift f{\"u}r Physik},
volume = {71},
pages = {205--226},
year = {1931},
doi = {10.1007/BF01341708}
}

@article{Valatin1961GHF,
author = {Valatin, J. G.},
title = {Generalized {Hartree-Fock} Method},
journal = {Physical Review},
volume = {122},
pages = {1012--1020},
year = {1961},
doi = {10.1103/PhysRev.122.1012}
}

@article{Bardeen1957Theory,
  author  = {Bardeen, J. and Cooper, L. N. and Schrieffer, J. R.},
  title   = {Theory of Superconductivity},
  journal = {Physical Review},
  volume  = {108},
  number  = {5},
  pages   = {1175--1204},
  year    = {1957},
  doi     = {10.1103/PhysRev.108.1175}
}

@article{Bogoliubov1958Superconductivity,
  author  = {Bogoliubov, N. N.},
  title   = {A New Method in the Theory of Superconductivity. I},
  journal = {Soviet Physics JETP},
  volume  = {7},
  pages   = {41--46},
  year    = {1958},
  note    = {[Zh. Eksp. Teor. Fiz. 34, 58 (1958)]}
}

@article{Valatin1958Superconductivity,
  author  = {Valatin, J. G.},
  title   = {Comments on the Theory of Superconductivity},
  journal = {Il Nuovo Cimento},
  volume  = {7},
  pages   = {843--857},
  year    = {1958},
  doi     = {10.1007/BF02745589}
}

@article{Schmid2004ProjectedHFB,
author = {Schmid, K. W.},
title = {On the use of general symmetry-projected {Hartree-Fock-Bogoliubov} configurations in variational approaches to the nuclear many-body problem},
journal = {Progress in Particle and Nuclear Physics},
volume = {52},
pages = {565--633},
year = {2004},
doi = {10.1016/j.ppnp.2004.02.001}
}

@article{JimenezHoyos2012ProjectedHF,
author = {Jim{\'e}nez-Hoyos, Carlos A. and Henderson, Thomas M. and Tsuchimochi, Takashi and Scuseria, Gustavo E.},
title = {Projected {Hartree-Fock} theory},
journal = {The Journal of Chemical Physics},
volume = {136},
pages = {164109},
year = {2012},
doi = {10.1063/1.4705280}
}

@article{Heisenberg1928,
  author  = {Heisenberg, W.},
  title   = {Zur Theorie des Ferromagnetismus},
  journal = {Zeitschrift f{\"u}r Physik},
  volume  = {49},
  pages   = {619--636},
  year    = {1928},
  doi     = {10.1007/BF01328601}
}

@article{Anderson1987,
  author  = {Anderson, P. W.},
  title   = {The Resonating Valence Bond State in {La$_2$CuO$_4$} and Superconductivity},
  journal = {Science},
  volume  = {235},
  number  = {4793},
  pages   = {1196--1198},
  year    = {1987},
  doi     = {10.1126/science.235.4793.1196}
}

@article{Tazhigulov2022,
  author  = {Tazhigulov, R. N. and Sun, S.-N. and Haghshenas, R. and Zhai, H. and Tan, A. T. K. and Rubin, N. C. and Babbush, R. and Minnich, A. J. and Chan, G. K.-L.},
  title   = {Simulating Models of Challenging Correlated Molecules and Materials on the {Sycamore} Quantum Processor},
  journal = {PRX Quantum},
  volume  = {3},
  number  = {4},
  pages   = {040318},
  year    = {2022},
  doi     = {10.1103/PRXQuantum.3.040318}
}

@article{Kitaev2006,
  author  = {Kitaev, A.},
  title   = {Anyons in an exactly solved model and beyond},
  journal = {Annals of Physics},
  volume  = {321},
  number  = {1},
  pages   = {2--111},
  year    = {2006},
  doi     = {10.1016/j.aop.2005.10.005}
}

@article{Whitfield2011QuantumSimulation,
  author  = {Whitfield, James D. and Biamonte, Jacob and Aspuru-Guzik, Al{\'a}n},
  title   = {Simulation of electronic structure Hamiltonians using quantum computers},
  journal = {Molecular Physics},
  volume  = {109},
  number  = {5},
  pages   = {735--750},
  year    = {2011},
  doi     = {10.1080/00268976.2011.552441}
}

@book{NielsenChuang2010QuantumComputation,
  author    = {Nielsen, Michael A. and Chuang, Isaac L.},
  title     = {Quantum Computation and Quantum Information},
  edition   = {10th Anniversary},
  publisher = {Cambridge University Press},
  address   = {Cambridge},
  year      = {2010}
}

@misc{ZhaoScuseriaDrudgeGristmill,
  author = {Zhao, Jinmo and Scuseria, Gustavo E.},
  title  = {{Drudge/Gristmill}},
  url    = {https://github.com/DrudgeCAS/drudge},
  note   = {Accessed: 2026-07-27}
}

@article{Johnson1988ModifiedBroyden,
author  = {Johnson, D. D.},
title   = {Modified {Broyden}'s method for accelerating convergence in self-consistent calculations},
journal = {Physical Review B},
volume  = {38},
pages   = {12807--12813},
year    = {1988},
doi     = {10.1103/PhysRevB.38.12807}
}

@article{Baran2008BroydenNuclear,
author  = {Baran, A. and Bulgac, A. and Forbes, M. M. and Hagen, G. and Nazarewicz, W. and Schunck, N. and Stoitsov, M. V.},
title   = {{Broyden}'s method in nuclear structure calculations},
journal = {Physical Review C},
volume  = {78},
pages   = {014318},
year    = {2008},
doi     = {10.1103/PhysRevC.78.014318}
}

@phdthesis{Gottesman1997Stabilizer,
  author = {Gottesman, Daniel},
  title  = {Stabilizer Codes and Quantum Error Correction},
  school = {California Institute of Technology},
  year   = {1997},
  eprint = {quant-ph/9705052},
  archivePrefix = {arXiv}
}

@article{Aaronson2004ImprovedStabilizer,
  author  = {Aaronson, Scott and Gottesman, Daniel},
  title   = {Improved simulation of stabilizer circuits},
  journal = {Physical Review A},
  volume  = {70},
  pages   = {052328},
  year    = {2004},
  doi     = {10.1103/PhysRevA.70.052328}
}

@article{Havlicek2017OperatorLocality,
  author  = {Havl{\'i}{\v c}ek, Vojt{\v e}ch and Troyer, Matthias and Whitfield, James D.},
  title   = {Operator locality in the quantum simulation of fermionic models},
  journal = {Physical Review A},
  volume  = {95},
  pages   = {032332},
  year    = {2017},
  doi     = {10.1103/PhysRevA.95.032332}
}

\end{document}